\documentclass[useAMS,usenatbib]{mn2e}
\usepackage{graphicx}
\usepackage{times}
\usepackage{amssymb, amsmath}
\usepackage{epstopdf}
\usepackage{color}
\usepackage[section]{placeins}
\usepackage[english]{babel}
\usepackage{lineno}
\usepackage{url}
\usepackage{ulem}
\def\mpchi{\,h^{-1}{\rm {Mpc}}}

\def\kpchi{\,h^{-1}{\rm {kpc}}}

\def\msun{\,h^{-1}{\rm M_{\sun}}}

\begin{document}

\title[Modeling the Luminosity and Colour Dependent Clustering]
{The clustering of galaxies in the SDSS-III Baryon Oscillation Spectroscopic
Survey: modeling of the luminosity and colour dependence in the Data Release
10}

\author[Guo et al.]{\parbox{\textwidth}{
Hong Guo$^{1,2}$\thanks{E-mail: hong.guo@utah.edu}, Zheng Zheng$^{1}$, Idit
Zehavi$^{2}$, Haojie Xu$^{1}$, Daniel J. Eisenstein$^{3}$, David H.
Weinberg$^{4,5}$, Neta A. Bahcall$^{6}$, Andreas A. Berlind$^{7}$, Johan
Comparat$^{8}$, Cameron K. McBride$^{3}$, Ashley J. Ross$^{9}$, Donald P.
Schneider$^{10,11}$, Ramin A. Skibba$^{12}$, Molly E. C. Swanson$^{3}$,
Jeremy L. Tinker$^{13}$, Rita Tojeiro$^{9}$, David A. Wake$^{14,15}$}
\vspace*{6pt} \\
$^{1}$ Department of Physics and Astronomy, University of Utah, UT 84112, USA\\
$^{2}$ Department of Astronomy, Case Western Reserve University, OH 44106, USA\\
$^{3}$ Harvard-Smithsonian Centre for Astrophysics, 60 Garden St., Cambridge, MA 02138, USA\\
$^{4}$ Department of Astronomy, Ohio State University, Columbus, OH 43210, USA\\
$^{5}$ Centre for Cosmology and Astro-Particle Physics, Ohio State University, Columbus, OH 43210, USA\\
$^{6}$ Department of Astrophysical Sciences, Princeton University, Peyton Hall, Princeton, NJ 08540, USA\\
$^{7}$ Department of Physics and Astronomy, Vanderbilt University, Nashville, TN 37235, USA\\
$^{8}$ Aix Marseille Universit\'e, CNRS, LAM (Laboratoire d'Astrophysique de
Marseille) UMR 7326, 13388, Marseille,
France\\
$^{9}$ Institute of Cosmology \& Gravitation, Dennis Sciama Building,
University of Portsmouth, Portsmouth, PO1 3FX,
UK\\
$^{10}$ Department of Astronomy and Astrophysics, The Pennsylvania State University, University Park, PA 16802, USA\\
$^{11}$  Institute for Gravitation and the Cosmos, The Pennsylvania State University, University Park, PA 16802, USA\\
$^{12}$ Centre for Astrophysics and Space Sciences, University of California,
9500 Gilman Drive, San Diego, CA 92093,
USA\\
$^{13}$ Centre for Cosmology and Particle Physics, New York University, New York, NY 10003, USA\\
$^{14}$ Department of Astronomy, University of Wisconsin-Madison, 475 N. Charter Street, Madison, WI, 53706, USA\\
$^{15}$ Department of Physical Sciences, The Open University, Milton Keynes
MK7 6AA, UK}

\maketitle

\begin{abstract}
We investigate the luminosity and colour dependence of clustering of CMASS
galaxies in the Sloan Digital Sky Survey-III Baryon Oscillation Spectroscopic
Survey Tenth Data Release, focusing on projected correlation functions of
well-defined samples extracted from the full catalog of $\sim540,000$
galaxies at $z\sim0.5$ covering about $6,500\,\deg^2$. The halo occupation
distribution framework is adopted to model the measurements on small and
intermediate scales (from $0.02$ to $60\mpchi$), infer the connection of
galaxies to dark matter halos and interpret the observed trends. We find that
luminous red galaxies in CMASS reside in massive halos of mass
$M{\sim}10^{13}$--$10^{14}\msun$ and more luminous galaxies are more
clustered and hosted by more massive halos. The strong small-scale clustering
requires a fraction of these galaxies to be satellites in massive halos, with
the fraction at the level of 5--8 per cent and decreasing with luminosity.
The characteristic mass of a halo hosting on average one satellite galaxy
above a luminosity threshold is about a factor $8.7$ larger than that of a
halo hosting a central galaxy above the same threshold. At a fixed
luminosity, progressively redder galaxies are more strongly clustered on
small scales, which can be explained by having a larger fraction of these
galaxies in the form of satellites in massive halos. Our clustering
measurements on scales below $0.4\mpchi$ allow us to study the small-scale
spatial distribution of satellites inside halos. While the clustering of
luminosity-threshold samples can be well described by a Navarro-Frenk-White
(NFW) profile, that of the reddest galaxies prefers a steeper or more
concentrated profile. Finally, we also use galaxy samples of constant number
density at different redshifts to study the evolution of luminous red
galaxies, and find the clustering to be consistent with passive evolution in
the redshift range of $0.5 \lesssim z \lesssim 0.6$.
\end{abstract}


\section{Introduction}
Galaxy luminosity and colour, the two readily measurable quantities, encode
important information about galaxy formation and evolution processes. The
clustering of galaxies as a function of luminosity and colour helps reveal
the role of environment in such processes. The dependence of clustering on
such galaxy properties is therefore a fundamental constraint on theories of
galaxy formation, and it is also important when attempting to constrain
cosmological parameters with galaxy redshift surveys,  since the different
types of galaxies trace the underlying dark matter distribution differently.
In this paper, we present the modeling of the luminosity and colour dependent
clustering of massive galaxies in the Sloan Digital Sky Survey-III
\citep[SDSS-III;][]{Eisenstein11}.

A fundamental measure of clustering is provided by measuring galaxy two-point
correlation functions (2PCFs). Galaxy clustering provides a powerful approach
to characterize the distribution of galaxies and probe the complex relation
between galaxies and dark matter. More luminous and redder galaxies are
generally observed, in various galaxy surveys, to have higher clustering
amplitudes than their fainter and bluer counterparts
\citep[e.g.][]{Davis76,Davis88,Hamilton88,Loveday95,Benoist96,Guzzo97,Norberg01,Norberg02,
Zehavi02,Zehavi05,Zehavi11,Budavari03,Madgwick03,Li06,Coil06,Coil08,Meneux06,Meneux08,Meneux09,
Wang07,Wake08,Wake11,Swanson08,Ross09,Ross10,Ross11,Skibba09,Skibba13,Loh10,Christodoulou12,Bahcall13,Guo13,Guo14}.

The clustering dependence of galaxies on their luminosity and colour can be
theoretically understood through the halo occupation distribution (HOD)
modeling \citep[see
e.g.][]{Jing98,Peacock00,Seljak00,Scoccimarro01,Berlind02,Berlind03,Zheng05,Zheng09,Miyatake13}
or the conditional luminosity function (CLF) method \citep{Yang03,Yang05}. In
HOD modeling, two determining factors that affect the clustering are the host
dark matter halo mass, $M$, and the satellite fraction $f_{\rm{sat}}$. The
emerging explanation for the observed trends is that more luminous galaxies
are generally located in more massive halos, while for galaxies of the same
luminosity, redder ones tend to have a higher fraction in the form of
satellite galaxies in massive halos \citep[e.g.][]{Zehavi11}. Residing in
more massive halos leads to a stronger clustering of galaxies on large
scales, while a higher satellite fraction results in stronger small-scale
clustering.

In this paper, we investigate the colour and luminosity dependent galaxy
clustering measured from the SDSS-III Baryon Oscillation Spectroscopic Survey
\citep[BOSS;][]{Dawson13} Data Release 10 \citep[DR10;][]{Anderson13,Ahn14}.
The SDSS-III BOSS survey is providing a large sample of luminous galaxies
that will allow a study of the galaxy-halo connection and the evolution of
massive galaxies (with a typical stellar mass of $10^{11.3}\msun$). By
carefully accounting for the effect of sample selections to construct nearly
complete subsamples, \cite{Guo13} (hereafter G13) investigated the luminosity
and colour dependence of galaxy 2PCFs from BOSS DR9 CMASS sample
\citep{Anderson12} in the redshift range of $0.43<z<0.7$. It was found that
more luminous and redder galaxies are generally more clustered, consistent
with the previous work. The evolution of galaxy clustering on large scales
(characterized by the bias factor) in the CMASS sample was also found to be
roughly consistent with passive evolution predictions.

While G13 presented the clustering measurements based on the DR9 sample, in
this paper we move a step forward to perform the HOD modeling to infer the
connection between galaxies and the hosting dark matter halos, using the
clustering measurements from the DR10 data. \cite{White11} presented the
first HOD modeling result for an early CMASS sample (from the first semester
of data). Now with DR10, the survey volume is more than 11 times larger than
that in \cite{White11}, which allows us to study the detailed relation
between the properties (specifically luminosity and colour) of the CMASS
galaxies and their dark matter halos. We build on similar studies for the
SDSS MAIN sample galaxies at $z{\sim}0.1$ \citep{Zehavi11} and luminous red
galaxies (LRGs) at $z{\sim}0.3$ \citep{Zheng09}, extending them now to higher
redshifts ($z{\sim}0.5$) and for galaxies at the high-mass end of the stellar
mass function. Given the key role of the CMASS galaxies as a large-scale
structure probe, it is also important to understand in detail how the CMASS
galaxies relate to the underlying dark matter halos for optimally utilizing
them for constraining cosmological parameters.

With about a factor of two increase in survey volume from DR9 to DR10, the
DR10 data produce more accurate measurements of the 2PCFs, and thus better
constraints on HOD parameters. After applying a fibre-collision correction,
with the method developed and tested in \cite*{Guo12}, we obtain good
measurements of the 2PCFs down to scales of ${\sim}20\kpchi$, with the help
of the larger survey area of DR10. This leads to the possibility of
determining the small-scale galaxy distribution profiles within halos, and we
also present the results of such a study.

The paper is organized as follows. In Section~\ref{sec:data}, we briefly
describe the CMASS DR10 sample and the clustering measurements for the
luminosity and colour samples. The HOD modeling method is presented in
Section~\ref{sec:hod}. We present our modeling results in
Section~\ref{sec:results} and give a summary in
Section~\ref{sec:conclusions}.

Throughout the paper, we assume a spatially flat $\Lambda$CDM cosmology (the
same as in G13), with $\Omega_m=0.274$, $h=0.7$, $\Omega_bh^2=0.0224$,
$n_s=0.95$, and $\sigma_8=0.8$.

\section{Data and Measurements}\label{sec:data}

\subsection{BOSS Galaxies and Luminosity and Colour Subsamples}

The SDSS-III BOSS selects galaxies for spectroscopic observations from the
five-band SDSS imaging data \citep{Fukugita96,Gunn98,Gunn06,York00}. A
detailed overview of the BOSS survey is given by \cite{Bolton12} and
\cite{Dawson13}, and the BOSS spectrograph is described in \cite{Smee13}.
BOSS is targeting $1.5$ million galaxies and $150,000$ quasars covering about
$10,000\deg^2$ of the SDSS imaging area. About $5$ per cent of the fibres are
devoted to more than $75,000$ ancillary targets probing a wide range of
different types of objects \citep{Dawson13}. In one BOSS ancillary program,
fibre-collided galaxies in the BOSS sample were fully observed in a small
area. We will present their clustering results in another paper (Guo et al.,
in preparation).

We focus on the analysis of the CMASS sample
\citep{Eisenstein11,Anderson12,Anderson13} selected from SDSS-III BOSS DR10.
The sample covers an effective area of about $6,500\deg^2$, almost twice as
large as in DR9. The selection of CMASS galaxies is designed to be roughly
stellar-mass limited at $z>0.4$. The detailed selection cuts are defined by,
\begin{eqnarray}
  17.5 < i_{\rm cmod} &<& 19.9 \\
  d_\perp &>& 0.55,\label{eq:dpcut} \\
  i_{\rm cmod} &<& 19.86 + 1.6(d_{\perp} - 0.8)\label{eq:slidecut} \\
  i_{\rm fib2} &<& 21.5 \\
  r_{\rm mod} - i_{\rm mod} &<& 2.0
\end{eqnarray}
where all magnitudes are Galactic-extinction corrected \citep*{Schlegel98}
and are in the observed frame. While the magnitudes are calculated using {\sc
cmodel} magnitudes (denoted by the subscript `cmod'), the colours are
computed using {\sc model} magnitudes (denoted by the subscript `mod'). The
magnitude $i_{\rm fib2}$ corresponds to the $i$-band flux within the fibre
aperture ($2^{\prime\prime}$ in diameter). The quantity $d_\perp$ in
Equations~(\ref{eq:dpcut}) and (\ref{eq:slidecut}) is defined as
\begin{equation}
  d_\perp = (r_{\rm mod} - i_{\rm mod}) - (g_{\rm mod} - r_{\rm mod})/8.
\end{equation}
Since the blue galaxies are generally far from complete in CMASS due to the
selection cuts of Equations (\ref{eq:dpcut}) and (\ref{eq:slidecut}) (see
also Figure~1 of G13), we focus in this paper on the clustering and evolution
of the red galaxies. The red galaxies in this paper are selected by a
luminosity-dependent colour cut (G13),
\begin{equation}
(r-i)>0.679-0.082(M_i+20)\label{eq:colourcut}
\end{equation}
where the absolute magnitude $M_i$ and $r-i$ colour are both k+e corrected to
$z=0.55$ \citep{Tojeiro12}.

\begin{table*}
\begin{minipage}{166mm}
\caption{Samples of different luminosity thresholds in the redshift range
$0.48<z<0.55$} \label{tab:lumsub}
\begin{tabular}{@{}llcccccccr@{}}
\hline
$M_i^{max}$ & $N_{\rm gal}$
            & $\bar{n}(z)$
            & $\chi^2/\rm{dof}$
            & $\log M_{\rm min}$
            & $\sigma_{\log M}$
            & $\log M_0$
            & $\log M_1^\prime$
            & $\alpha$
            & $f_{\rm{sat}}$(\rm{per\,cent})\\
\hline
$-21.6$     & 114417
            & $2.18\times10^{-4}$
            & 19.74/14
            & $13.37\pm0.05$
            & $0.58\pm0.05$
            & $0.57\pm2.09$
            & $14.30\pm0.02$
            & $1.56\pm0.03$
            & $7.91\pm0.43$\\
$-21.8$     & 65338
            & $1.25\times10^{-4}$
            & 28.83/14
            & $13.57\pm0.05$
            & $0.59\pm0.05$
            & $3.67\pm4.25$
            & $14.46\pm0.02$
            & $1.64\pm0.07$
            & $6.30\pm0.40$\\
$-22.0$     & 33964
            & $0.65\times10^{-4}$
            & 21.19/14
            & $13.80\pm0.06$
            & $0.61\pm0.06$
            & $2.81\pm3.07$
            & $14.59\pm0.03$
            & $1.82\pm0.09$
            & $5.04\pm0.36$\\
\hline
\end{tabular}
\medskip
The mean number density $\bar{n}(z)$ is in units of $h^{3}{\rm {Mpc}}^{-3}$.
The halo mass is in units of $\msun$. The satellite faction $f_{\rm{sat}}$ is
the derived parameter from the HOD fits. The best-fitting $\chi^2$ and the
degrees of freedom (dof) with the HOD modeling are also given. The degrees of
freedom are calculated as ${\rm dof}=N_{w_p}+1-N_{\rm par}$, where the total
number of data points ($N_{w_p}+1$) is that of the $w_p(r_p)$ data points
plus one number density data point, and $N_{\rm par}$ is the number of HOD
parameters.
\end{minipage}
\end{table*}

\begin{figure*}
\includegraphics[width=0.7\textwidth]{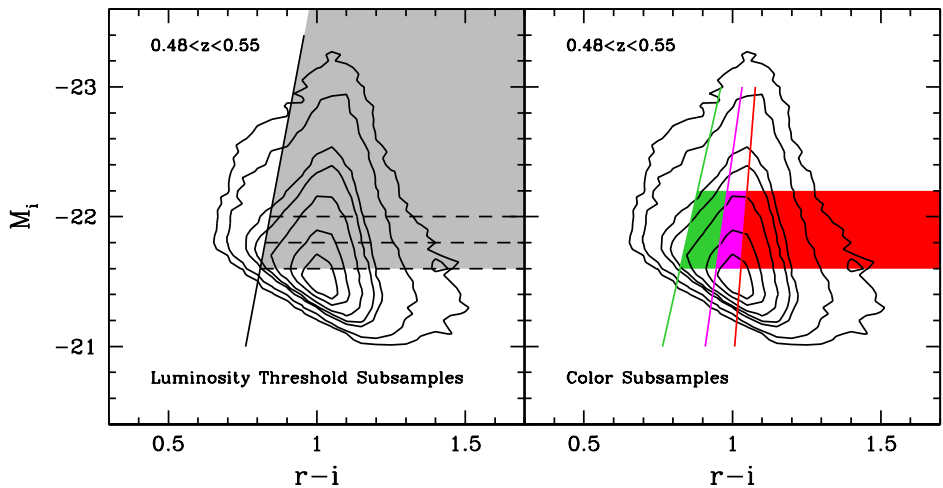}\caption{Left: the selection of the luminosity threshold samples in
the colour-magnitude diagram (CMD). The contours represent the density
distribution of the CMASS galaxies in the CMD. The shaded regions show the
galaxies covered in the luminosity threshold samples. The solid line denotes
the colour cut of Equation~(\ref{eq:colourcut}), and the three dashed lines
represent the three luminosity thresholds. Right: the selection of the `{\it
green}', `{\it redseq}' and `{\it reddest}' colour samples in the luminosity
range of $-22.2<M_i<-21.6$ and redshift range of $0.48<z<0.55$. The three
colour solid lines are the three cuts for the fine colour samples. The green
line is for the cut between the `{\it blue}' and `{\it green}' samples. The
magenta line is the cut between the `{\it green}' and `{\it redseq}' samples.
The red line is the cut between the `{\it redseq}' and `{\it reddest}'
samples. The shaded regions represent our selection of the corresponding
colour samples.} \label{fig:sample}
\end{figure*}
In this paper, we focus on modeling the luminosity and colour dependence of
the CMASS red galaxies. We therefore construct suitable luminosity and colour
subsamples of galaxies. Three luminosity threshold samples of red galaxies
are constructed, with $M_i<-21.6$, $M_i<-21.8$, and $M_i<-22.0$,  in the same
redshift range, $0.48<z<0.55$. We use luminosity threshold samples to
facilitate a more straight-forward HOD modeling of the measurements. The
redshift range is selected to ensure that the red galaxies in the samples are
nearly complete and minimally suffer from the selection effects (G13).
Details of the samples are given in Table~\ref{tab:lumsub} (together with the
best-fitting parameters from HOD modeling to be presented in
Section~\ref{sec:hod}). The left panel of Figure~\ref{fig:sample} shows the
selection of the three luminosity-threshold samples in the colour-magnitude
diagram (CMD). The contours represent the density distribution of the CMASS
galaxies in the CMD. The solid line denotes the colour cut of
Equation~(\ref{eq:colourcut}) and the shaded region and the three dashed
lines show the selection of the luminosity threshold samples.

For studying the colour dependence of clustering, we divide the CMASS
galaxies into `{\it green}', `{\it redseq}', and `{\it reddest}' subsamples,
using the colour cuts in Table~2 of G13. In order to have the best
signal-to-noise ratio and decouple the colour dependence from the luminosity
dependence, we only select galaxies in the redshift range of $0.48<z<0.55$
and luminosity range of $-22.2<M_i<-21.6$. The right panel in
Figure~\ref{fig:sample} shows the selection of the three subsamples. The
three coloured solid lines are the three cuts for the fine colour samples
(G13): the green, magenta, and red lines divide the galaxies into `{\it
blue}', `{\it green}', `{\it redseq}', and `{\it reddest}' subsamples,
respectively. The `{\it reddest}' sample has the reddest colour, while the
`{\it redseq}' sample represents the galaxies occupying the central part of
the red sequence in the CMD. The `{\it green}' sample is selected to
represent the transition from blue to red galaxies. The `{\it blue}' galaxy
sample is not considered here because of its low completeness. More
information on the colour subsamples can be found in Table~\ref{tab:colrsub}.

\subsection{Measurements of the Galaxy 2PCFs}

Approximately $1.5$ per cent of CMASS galaxies in DR10 were previously
observed in SDSS-II whose angular distribution differs from other BOSS
galaxies \cite[see more details in][]{Anderson12}. The redshift measurements
of these SDSS-II `Legacy' galaxies are, by construction, $100$ per cent
complete, while the redshift and angular completeness of BOSS galaxies vary
with sky position. The different distributions of these `Legacy' galaxies and
the newly observed BOSS galaxies need to be carefully taken into account for
clustering measurement. In previous work (e.g. \citealt{Anderson12},G13),
this is achieved by subsampling the SDSS-II galaxies to match the sector
completeness of BOSS survey. Here, to preserve the full information in the
`Legacy' galaxies, we adopt an alternative method, with a decomposition of
the total 3D correlation functions as follows (\citealt{Zu08}; Guo et al.
2012)
\begin{equation}
\xi_T={n_L^2\over n_T^2}\xi_{LL}+{2n_Ln_B\over n_T^2}\xi_{LB}+{n_B^2\over n_T^2}\xi_{BB}, \label{eq:xi_lb}
\end{equation}
where $n_L$, $n_B$ and $n_T$ are the number densities of the Legacy,
(uniquely) BOSS, and all galaxies, respectively, $\xi_{LL}$ is the
auto-correlation function of Legacy galaxies, $\xi_{BB}$ is the
auto-correlation function of BOSS galaxies, and $\xi_{LB}$ is the
cross-correlation of Legacy and BOSS galaxies. The decomposition can be
understood in terms of galaxy pair counts -- the total number of galaxy pairs
is composed of Legacy-Legacy pairs (related to $\xi_{LL}$), BOSS-BOSS pairs
(related to $\xi_{BB}$), and the Legacy-BOSS cross-pairs (related to
$\xi_{LB}$). The random samples are separately constructed for Legacy and
BOSS galaxies to reflect the different angular and redshift distributions.

In galaxy surveys using fibre-fed spectrographs, the precise small-scale
auto-correlation measurements are hindered by the effect that two fibres on
the same plate cannot be placed closer than certain angular scales, which is
$62^{\prime\prime}$ in SDSS-III, corresponding to about $0.4\mpchi$ at
$z{\sim}0.55$. Such fibre-collision effects can be corrected by using the
collided galaxies that are assigned fibres in the tile overlap regions, as
proposed and tested by Guo et al. (2012) and implemented in G13.

We apply the same method here to BOSS galaxies to correct for the
fibre-collision effect in measuring the 2PCFs. We divide the BOSS sample into
two distinct populations, one free of fibre collisions (labelled by subscript
`1') and the other consisting of potentially collided galaxies (labelled by
subscript `2'). With such a division, Equation~ (\ref{eq:xi_lb}) is further
decomposed into six terms as,
\begin{eqnarray}
\xi_T&=&{n_L^2\over n_T^2}\xi_{LL}+{2n_Ln_{B_1}\over n_T^2}\xi_{LB_1}+{2n_Ln_{B_2}\over n_T^2}\xi_{LB_2}\nonumber \\
&&+{n_{B_1}^2\over n_T^2}\xi_{B_1B_1}+{2n_{B_1}n_{B_2}\over n_T^2}\xi_{B_1B_2}+{n_{B_2}^2\over n_T^2}\xi_{B_2B_2}.
\end{eqnarray}
In actual measurements, the correlation functions $\xi_{LB_2}$,
$\xi_{B_1B_2}$, and $\xi_{B_2B_2}$ involving the collided galaxies are
estimated using the resolved collided galaxies in tile overlap regions (as
detailed in Guo et al. 2012).

We first measure the redshift-space 2PCF $\xi(r_p,r_\pi)$ in bins of
transverse separation $r_p$ and line-of-sight separation $r_\pi$ ($r_p$ in
logarithmic bins from $\sim$0.02 to $\sim$63$\mpchi$ with $\Delta\log
r_p=0.2$ and $r_\pi$ in linear bins from 0 to 100$\mpchi$ with $\Delta
r_\pi=2\mpchi$), using the \citet{Landy93} estimator. We then integrate the
2PCF along the line-of-sight direction to obtain the projected 2PCF
$w_p(r_p)$
\begin{equation}
w_p(r_p) = 2\int_0^{r_{\pi,\rm max}} \xi(r_p,r_\pi) dr_\pi,
\end{equation}
with $r_{\pi,\rm max}=100\mpchi$. This projected 2PCF is what we present and
model in this paper. The covariance error matrix for $w_p(r_p)$ is estimated
from 200 jackknife subsamples (\citealt{Zehavi02,Zehavi05},G13). We provide
the measurements for the projected 2PCF, $w_p(r_p)$, for all the subsamples
used in this paper in the Appendix~\ref{app:wp}. With the advantage of larger
sky coverage in DR10, the correlation function measurements have much smaller
errors compared with those in DR9. The measurements in DR9 and DR10 are
generally consistent within the errors.

\begin{table}
\caption{HOD parameters for the colour samples in $-22.2<M_i<-21.6$}
\label{tab:colrsub}
\begin{tabular}{@{}llcccr@{}}
\hline
Sample        & $N_{\rm gal}$
              & $\bar{n}(z)$
              & $\chi^2/\rm{dof}$
              & $\log M_1^\prime$
              & $f_{\rm{sat}}$(\rm{per\,cent})\\
\hline
`{\it green}'  & 28835
              & $0.54$
              & 20.81/16
              & $14.57\pm0.04$
              & $3.68\pm0.47$\\
`{\it redseq}' & 32221
              & $0.61$
              & 15.62/18
              & $14.42\pm0.03$
              & $6.15\pm0.46$\\
`{\it reddest}'& 34670
              & $0.66$
              & 35.26/18
              & $14.30\pm0.02$
              & $9.35\pm0.46$\\
\hline
\end{tabular}
\medskip
The mean number density $\bar{n}(z)$ is in unit of $10^{-4}h^{3}{\rm
{Mpc}}^{-3}$. The halo mass is in units of $\msun$. The redshift range of
colour samples are limited to $0.48<z<0.55$.
\end{table}

\section{HOD modeling}\label{sec:hod}
\begin{figure*}
\includegraphics[width=0.9\textwidth]{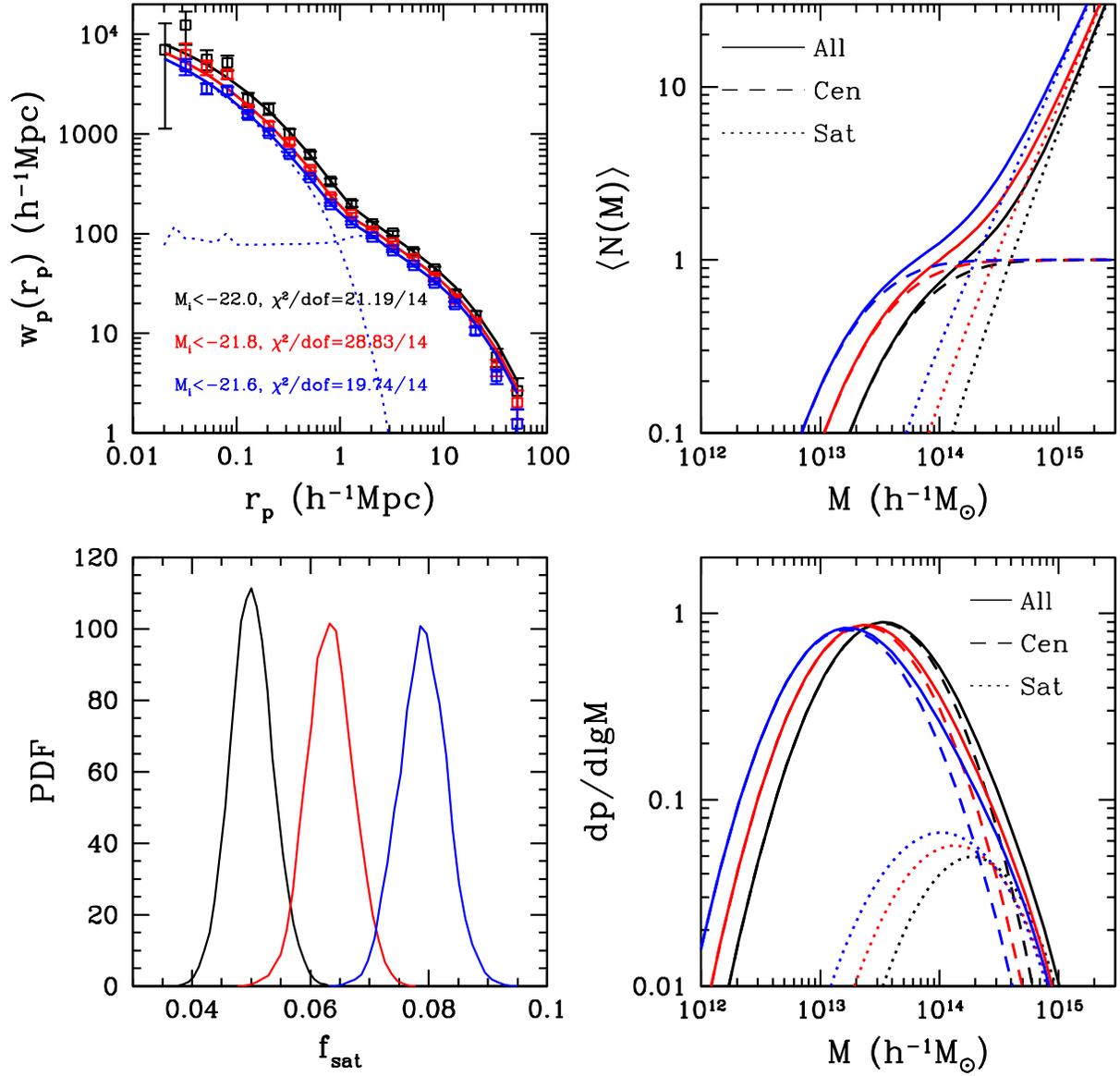}\caption{Projected two-point correlation functions measured in DR10
and the corresponding best-fitting HOD models for the three luminosity
threshold samples. Top left: the measurements of $w_p(r_p)$ from DR10
(squares) compared with the best-fitting HOD models (lines). The blue dotted
lines represent the one-halo and two-halo terms for the sample of
$M_i<-21.6$. The $\chi^2$ per degree of freedom (dof) for the three
best-fittings are also shown. Top right: the mean occupation number
distributions of the three samples. The total mean halo occupation function
(solid lines) is decomposed into contributions from central galaxies (dashed
lines) and satellite galaxies (dotted lines). Bottom left: The probability
distribution of $f_{\rm{sat}}$. Bottom Right: The probability distribution of
the host halo mass.} \label{fig:lum}
\end{figure*}
We perform the HOD fits to the projected two-point auto-correlation functions
$w_p(r_p)$, measured in different luminosity and colour bins. In the HOD
framework, it is helpful to separate the contribution to the mean number
$\langle N(M)\rangle$ of galaxies in halos of mass $M$ into those from
central and satellite galaxies \citep{Kravtsov04,Zheng05}.

For luminosity-threshold samples, we follow \cite*{Zheng07} to parameterize
the mean occupation functions of central and satellite galaxies as
\begin{eqnarray}
\langle N_{\rm cen}(M)\rangle&=&{1\over2}\left[1+{\rm erf}\left(\frac{\log M-\log M_{\rm min}}{\sigma_{\log
M}}\right)\right]\\
\langle N_{\rm sat}(M)\rangle&=&\langle N_{\rm cen}(M)\rangle\left(\frac{M-M_0}{M_1^\prime}\right)^\alpha
\end{eqnarray}
where ${\rm erf}$ is the error function. In total, there are five free
parameters in this parametrization. The parameter $M_{\rm min}$ describes the
cutoff mass scale of halos hosting central galaxies ($\langle N_{\rm
cen}(M_{\rm min})\rangle=0.5$). The cutoff profile is step-like but softened
to account for the scatter between galaxy luminosity and halo mass
\citep{Zheng05}, and is characterized by the width $\sigma_{\log M}$. The
three parameters for the mean occupation function of satellites are the
cutoff mass scale $M_0$, the normalization $M_1^\prime$, and the high-mass
end slope $\alpha$ of $\langle N_{\rm sat}(M)\rangle$. In halos of a given
mass, the occupation numbers of the central and satellite galaxies are
assumed to follow the nearest integer and Poisson distributions with the
above means, respectively. In our fiducial model, the spatial distribution of
satellite galaxies in halos is assumed to follow that of the dark matter,
i.e. the Navarro-Frenk-White (NFW) profile \citep*{Navarro97}, with halo
concentration parameter
\begin{equation}
c(M)=c_0(M/M_{\rm nl})^{\beta}(1+z)^{-1}, \label{eq:NFW}
\end{equation}
where $M_{\rm nl}$ is the non-linear mass scale at $z=0$, $c_0=11$ and
$\beta=-0.13$ \citep{Bullock01,Zhao09}. Later in this paper, we will also
consider a generalized NFW profile and use the 2PCF measurements to constrain
it. Halos here are defined to have a mean density 200 times that of the
background universe.

To theoretically compute the real-space 2PCF of galaxies within the HOD
framework, we follow the procedures laid out in \cite{Zheng04} and
\cite{Tinker05}. When computing the projected 2PCF from the real-space 2PCF,
we also incorporate the effect of residual redshift-space distortions to
improve the modeling on large scales. This is done by decomposing the 2PCF
into monopole, quadrupole, and hexadecapole moments and applying the method
of \citet{Kaiser87} (also see \citealt{Bosch13}).

We use a Markov Chain Monte Carlo (MCMC) method to explore the HOD parameter
space constrained by the projected 2PCF $w_p(r_p)$ and the number density
$n_g$ of each galaxy sample. The $\chi^2$ is formed as
\begin{equation}
\chi^2= \bmath{(w_p-w_p^*)^T C^{-1} (w_p-w_p^*)}
       +{(n_g-n_g^*)^2\over\sigma_{n_g}^2},
\end{equation}
where $\bmath{w_p}$ is the vector of $w_p$ at different values of $r_p$ and
$\bmath{C}$ is the full error covariance matrix determined from the jackknife
resampling method \citep[as detailed in][]{Guo13}. The measured values are
denoted with a superscript `$*$'. The error $\sigma_{n_g}$ on the number
density is determined from the variation of $n_g(z)$ in the different
jackknife subsamples. Finally, in order to account for the bias introduced
when inverting the covariance matrix \citep*{Hartlap07}, we multiply the
above $\chi^2$ by a factor $(n_{jk}-n_d-2)/(n_{jk}-1)$, which is about 0.9 in
our case. Here $n_{jk}$ is the number of jackknife samples and $n_d$ is the
dimension of the data vector. In Appendix~A, we demonstrate the robustness
and accuracy of our fitting with jackknife covariance matrices by comparing
with results from using mock covariance matrices.

\section{Modeling Results}\label{sec:results}
\subsection{HOD for the Luminosity-threshold Samples}
\begin{figure}
\includegraphics[width=0.42\textwidth]{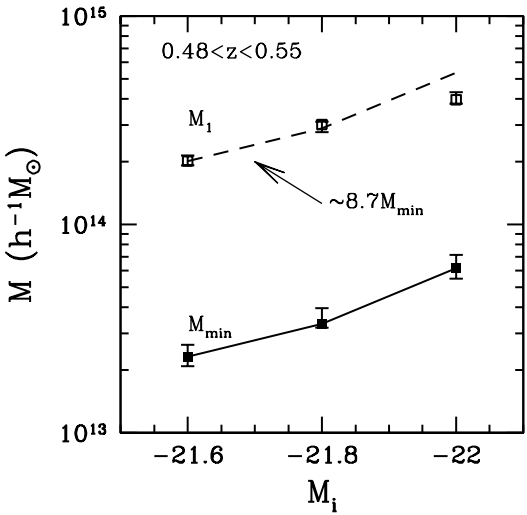}\caption{Two mass scales in HOD models as a function of threshold
luminosity of the galaxy samples, where $\langle N_{\rm cen}(M_{\rm min})\rangle=0.5$ and $\langle N_{\rm
sat}(M_1)\rangle=1$. The squares and solid lines are the HOD modeling results of the three luminosity threshold
samples,
while the dashed line shows the relation of $M_1{\sim}8.7M_{\rm min}$.} \label{fig:m1min}
\end{figure}
The best-fitting HOD parameters for the three luminosity threshold samples
are listed in Table~\ref{tab:lumsub}. Figure~\ref{fig:lum} shows the modeling
results. The top-left panel displays the measurements of $w_p(r_p)$ in DR10
(squares) compared with the best-fitting HOD models (lines). The top-right
panel shows the mean occupation functions from the three best-fitting models.
Overall, the trend of stronger clustering for more luminous CMASS samples is
explained in the HOD framework as a shift toward higher mass scale of host
halos, similar to that for the SDSS MAIN galaxies \citep{Zehavi11}.

The mean occupation functions in the top-right panel also show that a
fraction of the CMASS luminous red galaxies must be satellites in massive
halos. This is required to fit the small-scale clustering. The probability
distribution of $f_{\rm{sat}}$ is shown in the bottom-left panel of
Figure~\ref{fig:lum}. More luminous galaxies have a lower fraction of
satellites, consistent with the trend found for MAIN sample galaxies
\citep{Zehavi11} and luminous red galaxies \citep{Zheng09}. The peak
$f_{\rm{sat}}$ varies from $8$ per cent for the $M_i<-21.6$ sample to $5$ per
cent for the $M_i<-22.0$ sample.

The bottom-right panel of Figure~\ref{fig:lum} shows the probability
distributions of host halo mass for the three samples, generated from the
product of the mean occupation function and the differential halo mass
function \citep{Wake08,Zheng09}. The host halos refer to the main halos, i.e.
we do not consider the subhalos as the host halos. Most of the central
galaxies in these samples reside in halos of about a few times
$10^{13}\msun$, while the satellite galaxies are mostly found in halos of
masses around ${\sim}10^{14}\msun$. More luminous galaxies have a higher
probability to be found in more massive halos. For central galaxies, in the
narrow luminosity range in our samples, the peak host halo mass varies from
$1.1\times10^{13}$ to $3.3\times10^{13}\msun$ for the three samples.

\begin{figure*}
\includegraphics[width=0.9\textwidth]{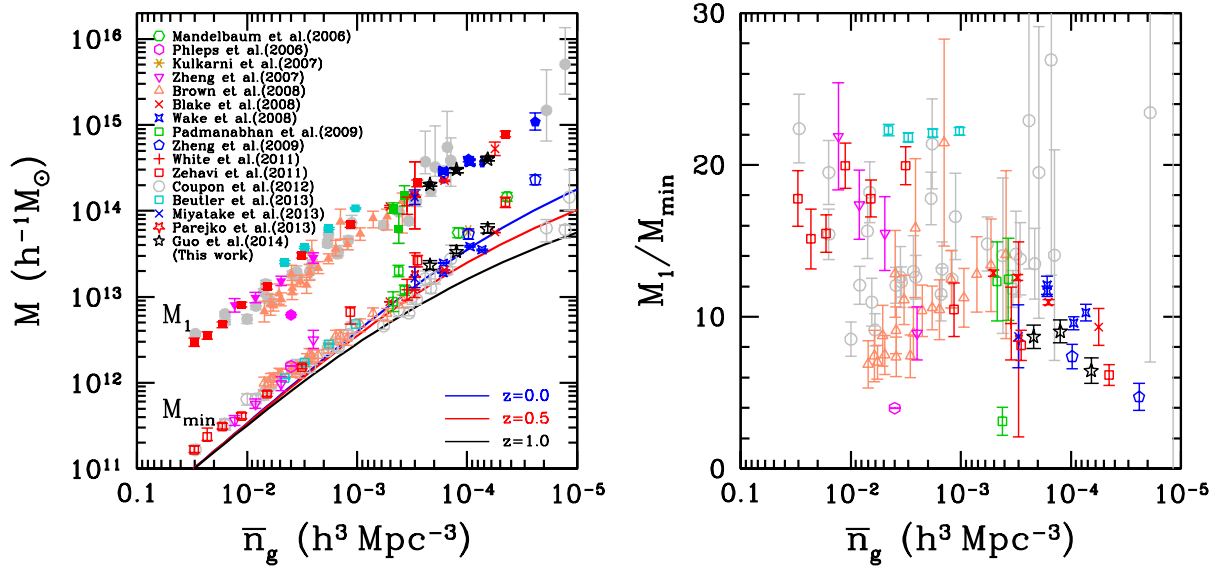}\caption{Left panel: HOD parameters $M_{\rm min}$ (open symbols) and
$M_1$ (solid symbols) as a function of the average number density $\bar{n}_g$
of the samples. Different symbols represent the measurements from the
literature, as labelled in the figure. Our measurements of the three
luminosity threshold samples are displayed by the black stars, which are in
good agreement with the literature. The halo mass functions at the three
typical redshifts $z=0$, $0.5$, and $1$ are also shown as the solid lines.
Right panel: the ratio between $M_1$ and $M_{\rm min}$ for all the
measurements in literature.} \label{fig:hodng}
\end{figure*}
Figure~\ref{fig:m1min} displays the relation of the HOD parameters $M_{\rm
min}$ and $M_1$ with the threshold luminosity $M_i$. Note that the quantity
$M_{\rm min}$ is the characteristic mass of halos hosting central galaxies at
the threshold luminosity (with $\langle N_{\rm cen}(M_{\rm
min})\rangle=0.5$), and $M_1$ is the characteristic mass of halos hosting on
average one satellite galaxy above the luminosity threshold ($\langle N_{\rm
sat}(M_1)\rangle=1$), which has a subtle difference from $M_1^\prime$.
Clearly, the tight correlation between galaxy luminosity and halo mass scales
persists for massive galaxies in massive halos at $z\sim 0.5$. The scaling
relation between $M_{\rm min}$ and $M_1$ in our samples roughly follows
$M_1{\sim}8.7M_{\rm min}$. The large gap between $M_1$ and $M_{\rm min}$
implies that a halo with mass between $M_{\rm min}$ and $M_1$ tends to host a
more massive central galaxy rather than multiple smaller galaxies
\citep{Berlind03}. This $M_1$-to-$M_{\rm min}$ ratio is comparable to the one
found for SDSS LRGs \citep{Zheng09} and significantly smaller than the
scaling factor found for the SDSS MAIN galaxies ($\sim17$,
\citealt{Zehavi11}). The ratio decreases somewhat with increasing luminosity
-- for the most luminous sample we analyse ($M_i<-22.0$), the ratio is
$6.4^{+0.5}_{-0.4}$, smaller than the ${\sim}8.7$ inferred from the lower
luminosity-threshold samples. Such a trend with luminosity is also found in
other SDSS analyses \citep{Zehavi05,Zehavi11,Skibba07,Zheng09}. These
behaviors are likely related to the dominance of accretion of satellites over
destruction in massive halos. More massive, cluster-sized halos form late and
accrete satellites more recently, leaving less time for satellites to merge
onto the central galaxies and thus lowering the satellite threshold mass
$M_1$ \citep{Zentner05}.

In Figure~\ref{fig:hodng} we compare the measurements of $M_{\rm min}$ and
$M_1$ of our samples with those from the literature of various surveys
\citep{Mandelbaum06,Phleps06,Kulkarni07,Zheng07,Brown08,Blake08,Wake08,
Padmanabhan09,Zheng09,White11,Zehavi11,Coupon12,Beutler13,Miyatake13,Parejko13}.
We plot them as a function of galaxy number density $\bar{n}_g$ (note that a
lower $\bar{n}_g$ corresponds to a higher threshold in galaxy luminosity or
stellar mass). The mass scales are all corrected to the cosmology adopted in
this paper according to their proportionality to $\Omega_m$
\citep[][]{Zheng02,Zheng09}. The left panel shows $M_{\rm min}$ (open
symbols) and $M_1$ (solid symbols) as a function of (decreasing) number
density of the different samples. The right panel displays the corresponding
ratios $M_1/M_{\rm min}$. Our results of the three luminosity threshold
samples (black stars) are in good agreement with the trend shown in other
samples.

\begin{figure}
\includegraphics[width=0.42\textwidth]{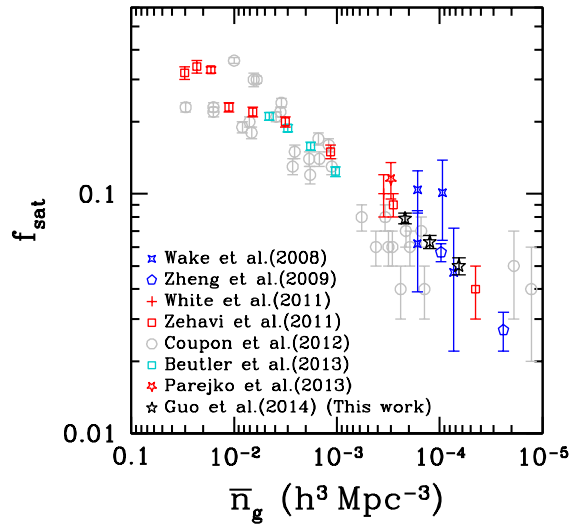} \caption{ \label{fig:fsat} Satellite fraction $f_{\rm{sat}}$ as a
function of the average number density $\bar{n}_g$ of the samples. Different symbols represent the measurements from
the
literature, as labelled in the figure. Our measurements of the three luminosity threshold samples are displayed by the
black stars.} \label{fig:fsatng}
\end{figure}

\cite{Brown08} noted that $M_{\rm min}$ and $\bar{n}_g$ approximately follow
a power-law relation with a power-law index ${\sim}-1$. Such a power-law
relation can be largely explained from the the halo mass function. In the
left panel, we plot the cumulative halo mass functions $n_h(>M)$ at three
typical redshifts, $z=0$, 0.5, and 1, as solid curves. The halo mass
functions are analytically computed for the assumed cosmological model. The
low-mass end ($M<10^{12}\msun$) of the halo mass function closely follows
$n_h(>M)\propto M^{-1}$ and evolves slowly with redshift. At the high-mass
end, the halo mass function drops more rapidly than the power-law at the low
mass end, and shows stronger redshift evolution. The halo mass function
$n_h(M>M_{\rm min})$ can be regarded as resulting from a simple form of HOD
--- one galaxy per halo and a sharp cutoff at $M_{\rm min}$, i.e. $\langle
N(M)\rangle=1$ for $M>M_{\rm min}$ and 0 otherwise, where $M_{\rm min}$ is
determined by matching the galaxy number density. Thus, any deviation from
the halo mass function curves could only be caused by the existence of
satellite galaxies and the softened mass cutoff around $M_{\rm min}$ for
central galaxies. For high number density galaxy samples, the deviation
arises from the satellite galaxies, since these halos have large satellite
fractions (see Figure~\ref{fig:fsat}). For low number density samples, the
prominent deviation is mainly a result of the wide softened cutoff in the
central galaxy mean occupation function. Such a wide softened cutoff is a
manifestation of the large scatter between central galaxy luminosity and halo
mass (Zheng et al. 2007). It is interesting that the deviations at both the
low- and high-mass end drive the $M_{\rm min}$--$\bar{n}_g$ relation towards
a power law.

The $M_1$--$\bar{n}_g$ relation also roughly follows a power law with a
slightly shallower slope than the $M_{\rm min}$--$\bar{n}_g$ relation. As a
consequence, there is a trend that the ratio $M_1/M_{\rm min}$ decreases with
decreasing $\bar{n}_g$, albeit with a large scatter, as shown in the right
panel of Figure~\ref{fig:hodng}. This result is consistent with what we find
in the luminosity dependence of $M_1/M_{\rm min}$.

Figure~\ref{fig:fsatng} presents the satellite fraction $f_{\rm{sat}}$ as a
function of the number density $\bar{n}_g$ from our luminosity-threshold
samples and those from the literature. The satellite fraction appears to
follow a well-defined sequence, especially toward low number density,
declining with decreasing number density and can be well described by a power
law, $f_{\rm sat}\simeq 0.1[\bar{n}_g/(10^{-3}h^3{\rm Mpc}^{-3})]^{1/3}$.
\begin{figure*}
\includegraphics[width=0.9\textwidth]{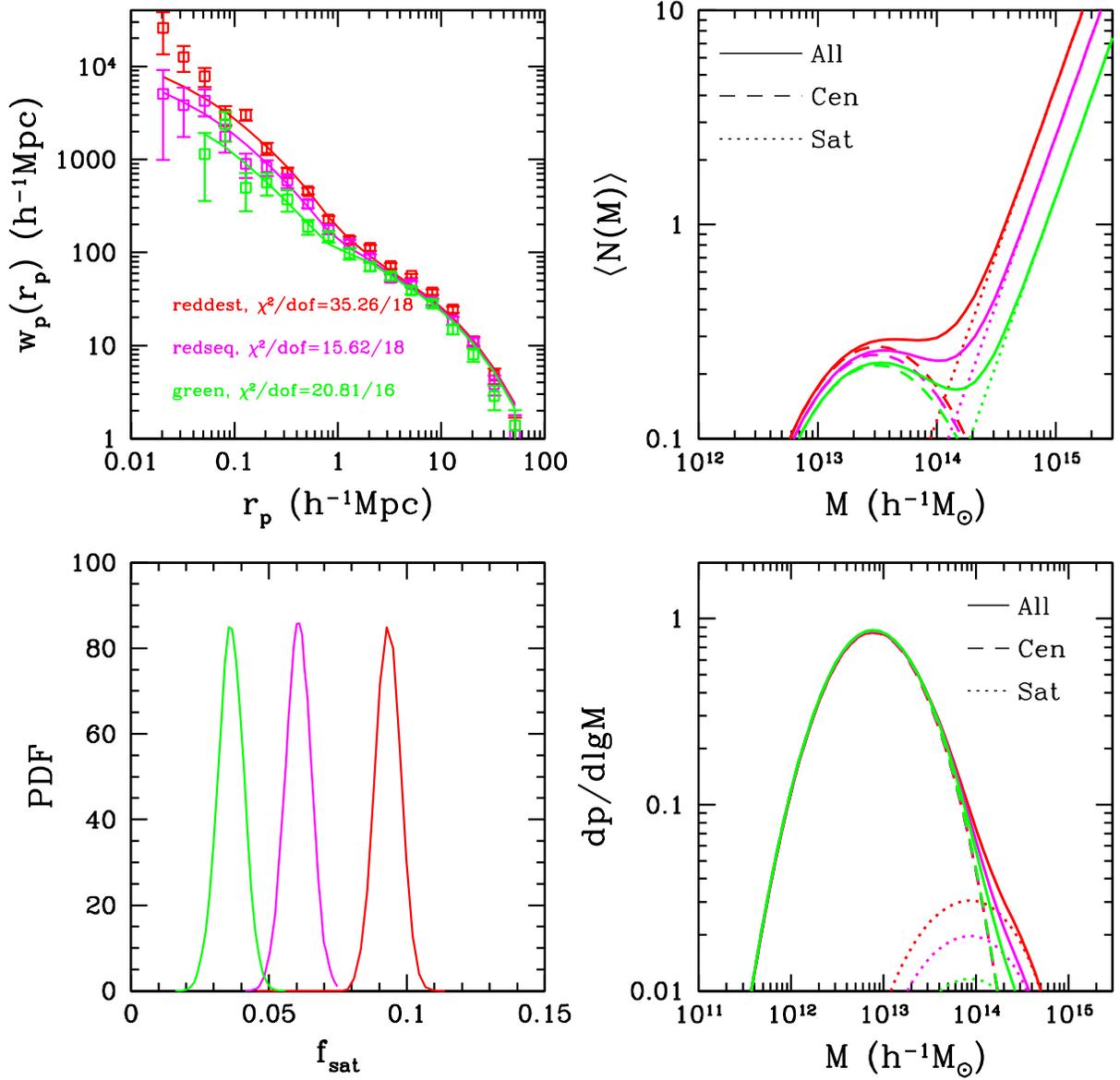}\caption{Similar to Figure~\ref{fig:lum}, but for the different
colour samples in the luminosity range $-22.2<M_i<-21.6$ and redshift range
$0.48<z<0.55$. The red, magenta, and green lines are for the `{\it reddest}',
`{\it redseq}' and `{\it green}' samples, respectively. } \label{fig:colour}
\end{figure*}
\subsection{HOD for the Colour Samples}\label{subsec:colour}

To model the colour dependence of the 2PCFs for the CMASS galaxies in the
luminosity bin of $-22.2<M_i<-21.6$, we form the mean occupation function of
the central galaxies in this luminosity bin as the difference between
$\langle N_{\rm cen}(M)\rangle$ of the $M_i<-21.6$ sample and that of the
$M_i<-22.2$ sample. Following \cite{Zehavi11}, we fix the slope $\alpha$ of
the satellite mean occupation function to be 1.56, the value from the fainter
luminosity threshold sample that dominates the number density of the
luminosity-bin sample. The cutoff mass of the mean occupation function of
satellite galaxies is also set to be the smaller of the two values from the
two threshold samples. Thus, we model the 2PCF of each colour subsample with
only one free parameter, $M_1^\prime$. In this simple model, the shape of the
central or satellite mean occupation functions for different colour samples
remains the same, and the relative normalization between the central and
satellite mean occupation functions is governed by $M_1^\prime$ and
constrained by the small-scale clustering. The overall normalizations of the
mean occupation functions are determined from the relative number densities
of the colour samples to the total number density in this luminosity bin. By
construction, the sum of the mean galaxy occupation functions of all colour
samples (including the `{\it blue}' galaxies that are not modeled in this
paper) equals that of the full luminosity-bin sample.

Figure~\ref{fig:colour} shows the modeling result (also in
Table~\ref{tab:colrsub}) of the three colour samples, in a similar format to
that of Figure~\ref{fig:lum}. It is evident from the figure that redder
galaxies have a higher clustering amplitude, especially on small scales
(one-halo term). Within our model, the higher clustering amplitude in the
redder galaxies is a result of a larger fraction of them being satellites (in
massive halos), as shown in the top-right and bottom-left panels. The
satellite fraction $f_{\rm sat}$ varies from $3.7$ per cent in the `{\it
green}' sample to $9.4$ per cent in the `{\it reddest}' sample, consistent
with the trend in MAIN sample \citep{Zehavi05,Zehavi11}. Compared with the
luminosity dependence of galaxy clustering, where more luminous galaxies
reside in more massive halos and have smaller satellite fractions, the trend
in the colour dependence indicates that the satellite fraction mostly affects
the small-scale clustering, while halo mass scales affect the overall
clustering amplitudes (especially on large scales dominated by the two-halo
term).

The top-left panel of Figure~\ref{fig:colour} demonstrates that the projected
2PCF of the `{\it reddest}' sample on small scales ($r_p<0.2\mpchi$) is not
well fitted by the simple HOD model, leading to a high value of best-fitting
$\chi^2$. Even if we allow the slope $\alpha$ of the mean satellite
occupation function to vary, the situation does not significantly improve. We
explore the implication for the small-scale galaxy distribution inside halos
with a generalized NFW profile in Section~\ref{subsec:gnfw}.

A close inspection of the best-fitting projected 2PCFs in the top-left panel
shows that the model predicts a narrower range of clustering amplitude than
that from the data on scales above a few $\mpchi$. Since the large-scale
amplitude in the 2PCF is mainly determined by central galaxies, this result
implies that the halo mass scales for central galaxies in the luminosity bin
can vary with the colour to some degree (in the sense of higher mass scales
for redder central galaxies) in a manner that is not captured in our simple
model.

\subsection{Generalized NFW profile}\label{subsec:gnfw}
\begin{figure*}
\includegraphics[width=0.9\textwidth]{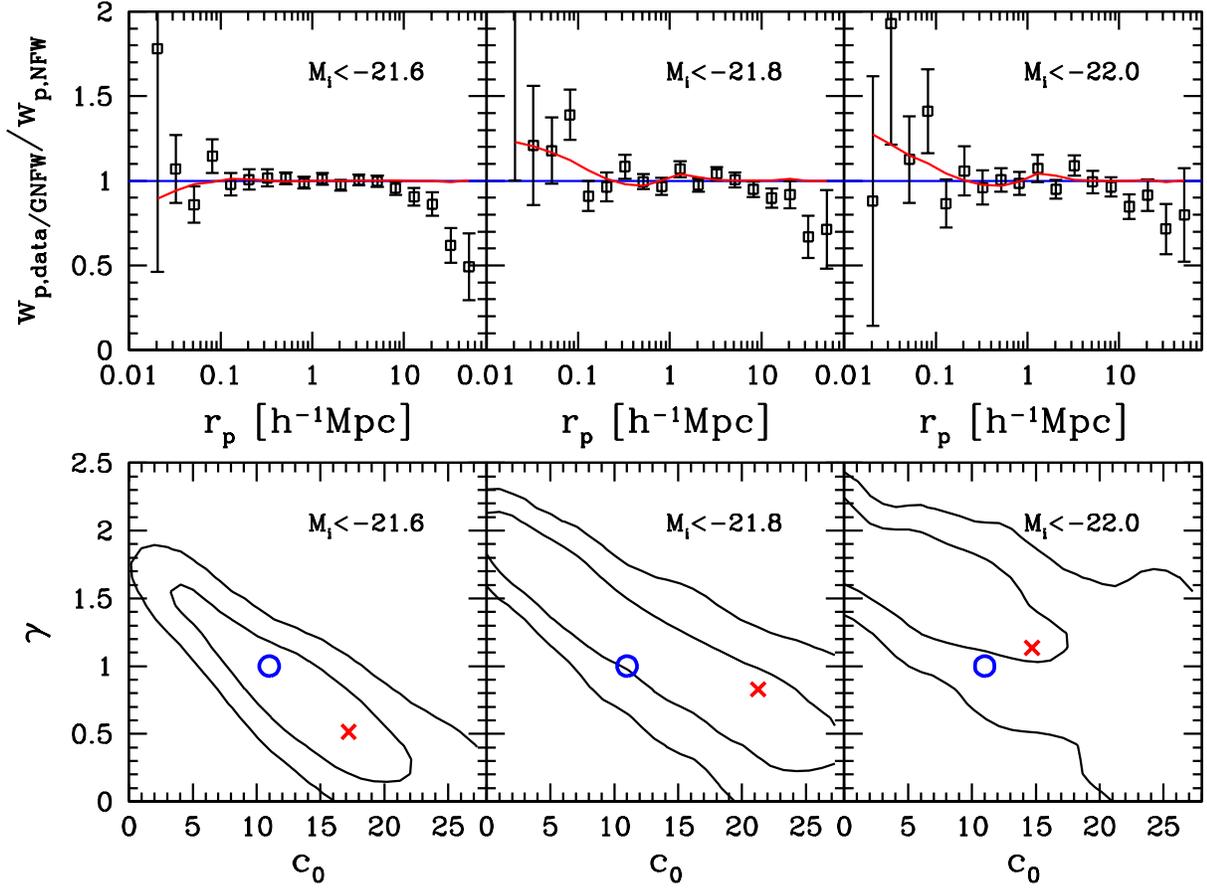}\caption{Top panels: ratios of $w_p(r_p)$ predicted from the
best-fitting GNFW HOD models and those measured from the data to those of the
NFW model predictions. The black squares represent the data measurements,
while the red lines are the predictions of the GNFW models. Bottom panels:
marginalized joint distribution of the concentration parameter $c_0$ and the
slope $\gamma$ in the generalized NFW model for the three luminosity
threshold samples. The contours show the $68$ per cent and $95$ per cent
confidence levels for the two parameters. The red crosses represent the
best-fitting GNFW HOD models, while the blue circles are the predictions of
the NFW model.} \label{fig:cgamma_lum}
\end{figure*}
In our fiducial HOD model, we assume that the spatial distribution of
satellite galaxies inside halos follows the same NFW profile as the dark
matter. The best-fitting small-scale clustering amplitude for the `{\it
reddest}' sample shows deviations from the data (see
Figure~\ref{fig:colour}), implying a possible departure of the satellite
distribution from the NFW profile. To explore such a possibility, we also
consider a generalized NFW (hereafter GNFW) profile to describe the
distribution of satellite galaxies inside halos by allowing two more free
parameters in the HOD model, the normalization parameter $c_0$ for the halo
concentration in Equation~(\ref{eq:NFW}) and the slope $\gamma$ in the
density profile \citep{Watson10,Watson12,Bosch13},
\begin{equation}
\rho(r)\propto \left[\left({c\,r\over r_{\rm vir}}\right)^\gamma\left(1+{c\,r\over r_{\rm
vir}}\right)^{3-\gamma}\right]^{-1}
\end{equation}
where $r_{\rm vir}$ is the virial radius of the halo. As a special case, the
NFW profile has $c_0=11$ (Equation \ref{eq:NFW}) and $\gamma=1$. Another
special case is the singular isothermal sphere (SIS) distribution, which has
$c_0\rightarrow0$ and $\gamma=2$.

\begin{table}
\caption{$\chi^2/\rm{dof}$ in generalized NFW models} \label{tab:hodgnfw}
\begin{tabular}{@{}lcccc@{}}
\hline
Sample          & $\chi^2/\rm{dof}$
                & $\Delta\rm{AIC}$
                & $\exp(\Delta\rm{AIC}/2)$
                & $f_{\rm{sat}} (\rm{per\,cent})$\\
\hline
$M_i<-21.6$     & $19.34/12$
                & 3.60 & 6.05
                & $7.76\pm0.55$\\
$M_i<-21.8$     & $24.75/12$
                & -0.08
                & 0.96
                & $5.77\pm0.55$\\
$M_i<-22.0$     & $18.93/12$
                & 1.74
                & 2.39
                & $4.66\pm0.56$\\
\hline
`{\it green}'    & 17.84/14
                & 1.03
                & 1.67
                & $4.11\pm0.53$\\
`{\it redseq}'   & 11.76/16
                & 0.14
                & 1.07
                & $6.50\pm0.61$\\
`{\it reddest}'  & 28.94/16
                & -2.32
                & 0.31
                & $9.73\pm0.62$\\
\hline
\end{tabular}
\medskip
$\Delta\rm{AIC}=\rm{AIC_{GNFW}}-\rm{AIC_{NFW}}$
\end{table}
We first apply the GNFW model to the luminosity-threshold samples. From
Figure~\ref{fig:lum}, the HOD model using the NFW profile can fit the 2PCFs
of the luminosity-threshold samples reasonably well. By including the two
additional free parameters, the best-fitting $\chi^2$ values only decrease
slightly, as shown in Table~\ref{tab:hodgnfw}. To compare the goodness of
fits between the generalized and original NFW profiles, we make use of the
Akaike information criterion \cite[AIC;][]{Akaike74}, defined as ${\rm
AIC}=\chi^2+2k$ for each model, where $k$ is the number of HOD parameters.
The difference $\Delta {\rm AIC}\equiv\rm{AIC_{GNFW}}-\rm{AIC_{NFW}}$ between
the AIC values of the GNFW and NFW models reveals that the model with the NFW
profile is $\exp(\Delta\rm{AIC}/2)$ times as probable as that with the GNFW
profile. As shown in Table~\ref{tab:hodgnfw}, only the $M_i<-21.8$ sample
shows a marginal preference for the GNFW profile. Overall, the clustering
data of the luminosity-threshold samples do not require a profile different
than the NFW profile.

\begin{figure}
\includegraphics[width=0.42\textwidth]{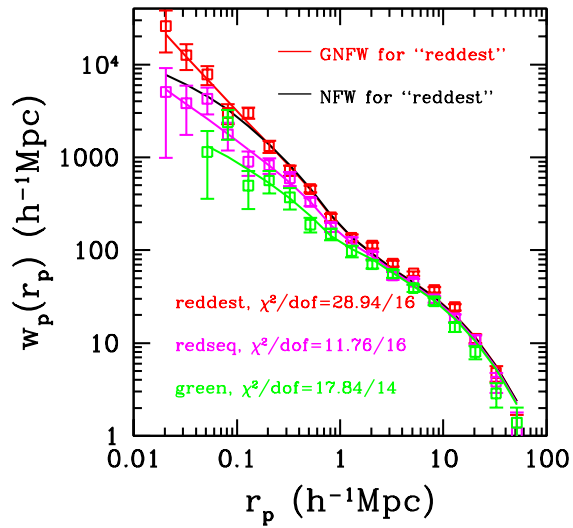} \caption[]{The same measurements of Figure~\ref{fig:colour},
but now with the GNFW best-fitting models. The red, magenta, and green lines are for the `{\it reddest}', `{\it
redseq}'
and `{\it green}' samples, respectively. For comparison, we also show the best-fitting NFW model for the `{\it
reddest}'
sample as the black line.} \label{fig:colour_GNFW}
\end{figure}
To present a detailed examination of the effect of the GNFW profile, we show
in the top panels of Figure~\ref{fig:cgamma_lum} the ratios of $w_p(r_p)$
predicted from the best-fitting GNFW HOD models (red lines) and the measured
$w_p(r_p)$ (squares) to those of the NFW model predictions. Both the NFW and
GNFW models fit the data reasonably well on scales of $r_p<10\mpchi$. Their
predictions are similar on large scales and they only differ slightly on
scales below $\sim$1$\mpchi$. On larger scales, the models appear to
overestimate the large scale bias for all samples, which might be caused by
the sample variance.

\begin{figure*}
\includegraphics[width=0.9\textwidth]{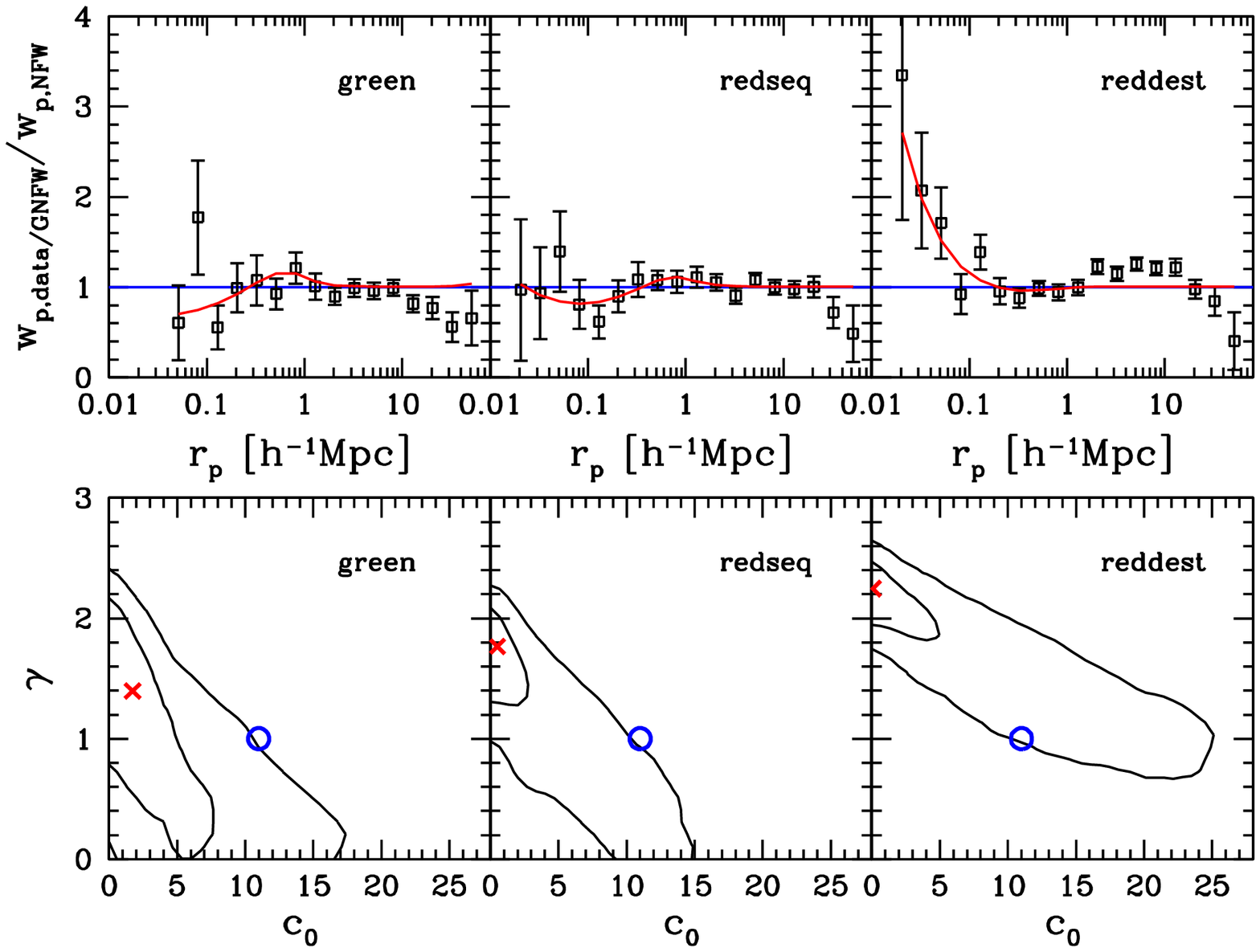}\caption{Similar to Figure~\ref{fig:cgamma_lum}, but for the
three colour samples in the luminosity bin $-22.2<M_i<-21.6$.}
\label{fig:cgamma_colour}
\end{figure*}
The marginalized joint distribution of the concentration normalization $c_0$
and the slope $\gamma$ for the three luminosity threshold samples are shown
in the bottom panels of Figure~\ref{fig:cgamma_lum}. The best-fitting models
are displayed as the red crosses and the NFW model is represented by the blue
circles. While there is a weak trend that the profile for more luminous
samples prefers to deviate from the NFW profile, the NFW profile is still
within the $\sim 2\sigma$ range of the contours. \citet{Watson10} and
\cite{Watson12} find that the distributions of satellite LRGs and satellites
of luminous galaxies in SDSS MAIN sample have significantly steeper inner
slopes than the NFW profile. From Figure~1 in both \citet{Watson10} and
\cite{Watson12}, we infer that constraining the deviation from the NFW
profile requires accurate measurements on scales $r_p<0.03\mpchi$. However,
at such small scales, the effect of photometric blending of close pairs may
be important in clustering measurements \citep{Masjedi06,Jiang12}, which is
not corrected for in our samples. Moreover, the measurement errors at these
scales are large in our samples. We therefore can only conclude that our
measurements and modeling results show no strong deviations from the NFW
profile in the distribution of satellites inside halos for
luminosity-threshold samples.

We then apply the GNFW model to the fine colour samples in
Section~\ref{subsec:colour}. Significant improvement over the NFW profile
model is found in fitting the small-scale 2PCF of the `{\it reddest}' sample,
as shown in Figure~\ref{fig:colour_GNFW}. Without the variation in $c_0$ and
$\gamma$, the pure NFW model cannot fit well the small-scale clustering by
only adjusting the satellite fraction or the slope $\alpha$ of the satellite
mean occupation function. The best-fitting $\chi^2$ value for this colour
sample is also reduced by adding the two free parameters, as shown in
Table~\ref{tab:hodgnfw}. From the difference in the AIC, the NFW model is
much less favorable than the GNFW model for the `{\it reddest}' sample, while
it still provides reasonable fits to the `{\it green}' and `{\it redseq}'
samples.

The ratios of the HOD model predictions of GNFW models and the measured
$w_p(r_p)$ to those of the NFW model are shown in the top panels of
Figure~\ref{fig:cgamma_colour}. The small-scale clustering of the `{\it
reddest}' sample is better fit by the GNFW model, which confirms the
importance of the small-scale ($r_p<0.1\mpchi$) measurements in
distinguishing the NFW and GNFW models. The marginalized joint distribution
of $c_0$ and $\gamma$ is presented in the bottom panels of
Figure~\ref{fig:cgamma_colour}. From the $1\sigma$ contours of
Figure~\ref{fig:cgamma_colour}, there is a visible trend that redder galaxies
favor smaller $c_0$ and larger $\gamma$. The SIS profile (corresponding to
$c_0\rightarrow0$ and $\gamma=2$) seems to provide better fits than the NFW
profile, consistent with the previous findings \citep{Grillo12,Watson12}.
This trend is clearly manifested for the `{\it reddest}' sample.

\begin{figure}
\includegraphics[width=0.42\textwidth]{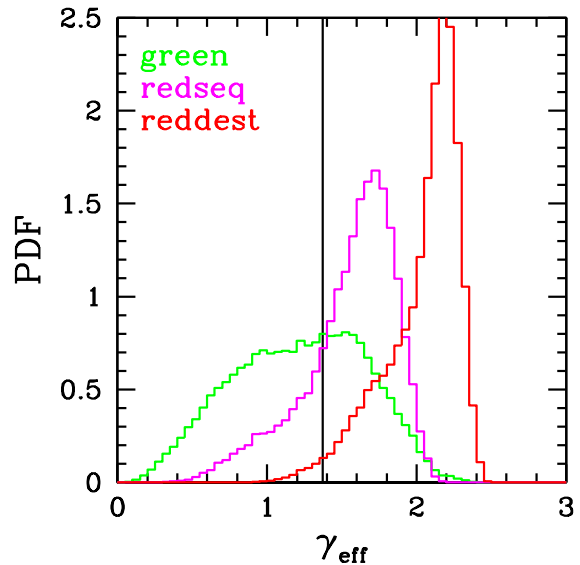}\caption{Probability distributions of the effective slope $\gamma_{\rm
eff}$ at $r=0.1\mpchi$ in halos of $2\times10^{14}\msun$ for the three colour samples.
The vertical black line denotes $\gamma_{\rm eff}=1.37$, the value for the NFW
profile at this radius.} \label{fig:reff}
\end{figure}
Since $c_0$ and $\gamma$ are correlated, different combinations of them can
lead to similar shape of the profile. A better quantity to represent the
shape of the density profile is the effective slope defined as
\begin{equation}
\gamma_{\rm eff}\equiv-\frac{d\ln\rho(r)}{d\ln r}=\gamma+\frac{c(3-\gamma)}{c+r_{\rm vir}/r}.
\end{equation}
It is the local slope of the profile at radius $r$ for halos of virial radius
$r_{\rm vir}$. That is, if approximated by a power law, the local profile is
proportional to $r^{-\gamma_{\rm eff}}$. To understand the trend shown in
Figure~\ref{fig:cgamma_colour} for the three colour samples, we show in
Figure~\ref{fig:reff} the probability distributions of the effective slope
$\gamma_{\rm eff}$ at $r=0.1\mpchi$ in halos of $2\times10^{14}\msun$. For
reference, the vertical black line denotes the $\gamma_{\rm eff}$ for the NFW
profile at the same radius. There is a clear trend that the distribution of
redder galaxies has a steeper effective slope, as expected from the top
panels of Figure~\ref{fig:cgamma_colour}. While the effective slopes for the
`{\it green}' and `{\it redseq}' galaxies still agree with the NFW model,
that for the `{\it reddest}' colour sample deviates significantly from the
NFW value. The above model assumes a fixed slope $\alpha$ for $\langle N_{\rm
sat}\rangle$. If we allow $\alpha$ to vary, each probability distribution
curve in Figure~\ref{fig:reff} becomes broader (as well as the range given by
the contours in Figure~\ref{fig:cgamma_colour}), but the trend remains the
same and the result is still valid. We therefore conclude that the steep rise
in the small-scale 2PCF of the `{\it reddest}' colour sample favors a GNFW
profile with a steeper inner slope.

\subsection{Central-Satellite Correlation}
\begin{figure*}
\includegraphics[width=0.9\textwidth]{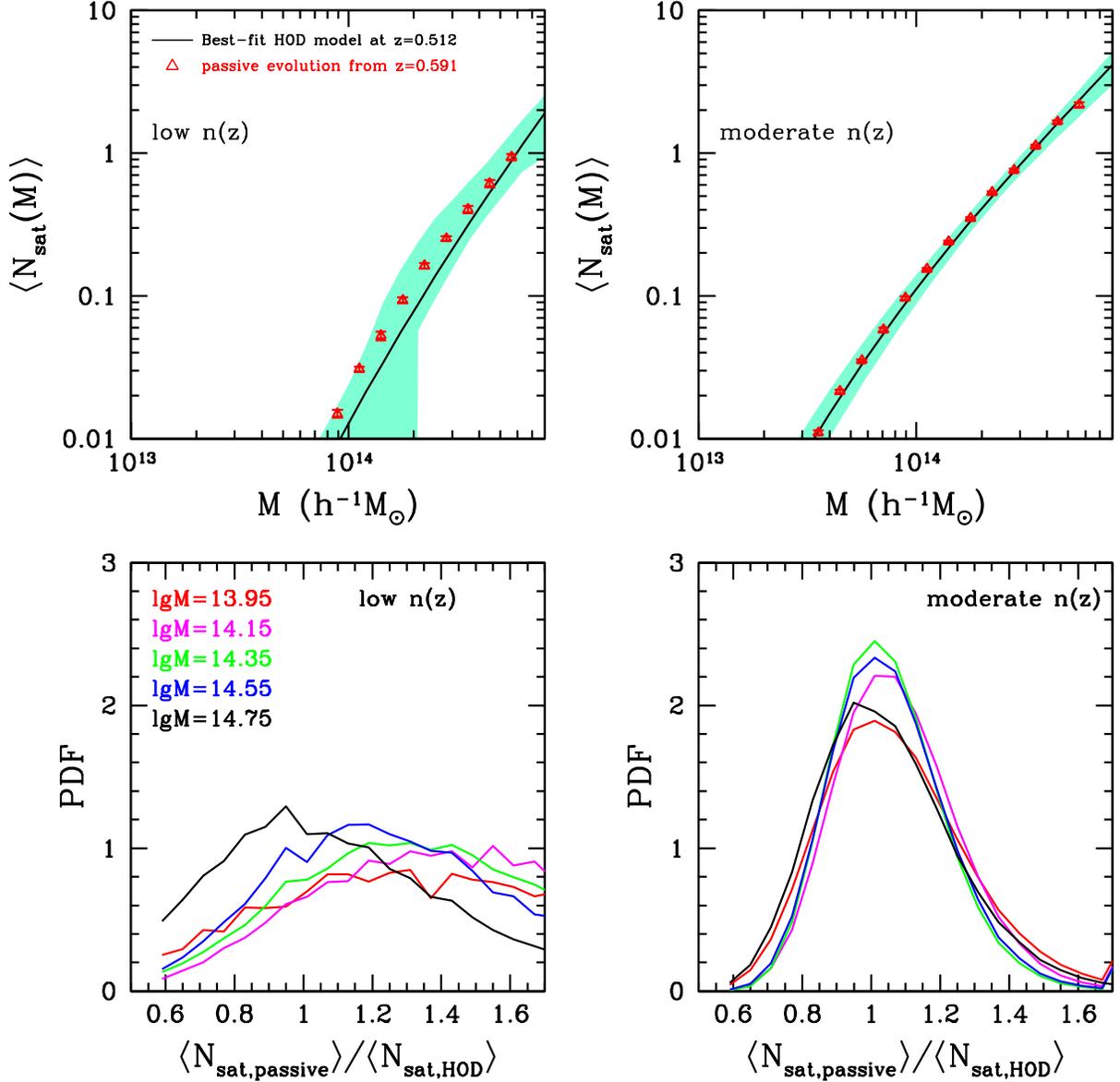}\caption{Top panels: satellite occupation distributions for the low
and moderate $n(z)$ samples. The solid lines are the best-fitting HOD models
at $z=0.512$ and the green shaded areas are the $1\sigma$ distribution. The
triangles are the mean occupation function obtained by using the simulations
to passively evolve the best-fitting $z=0.591$ HOD to $z=0.512$. Bottom
panels: probability distributions of the ratios between the satellite
occupation numbers of the passively evolved HOD and the best-fitting HOD.
Different colour lines correspond to different host halo masses.}
\label{fig:passive}
\end{figure*}
For the HOD model in previous sections, we make an implicit and subtle
assumption in the one-halo central-satellite galaxy pairs, related to the
correlation between central and satellite galaxies. For the contribution of
the one-halo central-satellite galaxy pairs in the model, one needs to
specify the mean $\langle N_{\rm cen} N_{\rm sat}\rangle$ at each halo mass.
We compute it as $\langle N_{\rm cen} N_{\rm sat}\rangle = \langle N_{\rm
sat} \rangle$. This result implies that for a given galaxy sample, a halo
hosting a satellite galaxy also hosts a central galaxy from the same sample
\citep[][]{Zheng05,Simon09}. Such an assumption is quite reasonable for
luminosity-threshold samples, if the central galaxy is the most luminous
galaxy in a halo. However, for a luminosity-bin sample or colour subsamples,
a scenario can arise where the central galaxy in a halo hosting satellite
galaxies from the sample does not itself belong to the sample.

To explore the effect of the central-satellite correlation, we consider an
extreme case in which the occupations of central and satellite galaxies
inside halos are completely independent, i.e. the probability of a halo to
host a satellite does not depend on whether it has a central galaxy from the
same sample. The mean number of central-satellite pairs is then computed as
$\langle N_{\rm cen} N_{\rm sat}\rangle = \langle N_{\rm cen}\rangle \langle
N_{\rm sat} \rangle$.

\begin{table}
 \centering
\begin{minipage}{86mm}
\caption{HOD Model Fitting Results for independent Central-Satellite
Distribution} \label{tab:hodcscorr}
\begin{tabular}{@{}lcccc@{}}
\hline
Sample             & $({\chi^2\over\rm{dof}})_{\rm{NFW}}$
                   & $({\chi^2\over\rm{dof}})_{\rm{GNFW}}$
                   & $f_{\rm{sat,NFW}}$
                   & $f_{\rm{sat,GNFW}}$\\
\hline
$M_i<$-21.6        & $21.89/14$
                   & $19.38/12$
                   & $9.92\pm0.55$
                   & $8.88\pm0.73$\\
$M_i<$-21.8        & $33.02/14$
                   & $25.14/12$
                   & $8.26\pm0.59$
                   & $6.93\pm0.78$\\
$M_i<$-22.0        & $22.56/14$
                   & $18.69/12$
                   & $6.95\pm0.71$
                   & $3.98\pm1.17$\\
\hline
`{\it green}'       & 22.17/16
                   & 21.59/14
                   & $7.02\pm0.50$
                   & $6.90\pm0.73$\\
`{\it redseq}'      & 14.52/18
                   & 12.88/16
                   & $10.14\pm0.60$
                   & $10.03\pm0.59$\\
`{\it reddest}'     & 52.26/18
                   & 24.26/16
                   & $12.98\pm0.52$
                   & $13.20\pm0.50$\\
\hline
\end{tabular}
\medskip
$f_{\rm{sat}}$ is in unit of per cent.
\end{minipage}
\end{table}
Table~\ref{tab:hodcscorr} summarizes the HOD modeling results for independent
central-satellite occupations. The general results and trends stay similar to
those previously discussed. Compared with the fitting results obtained in the
previous sections, the satellite fractions are somewhat increased, with the
increase for the colour samples more substantial (a factor of 1.4--1.9). The
change in satellite fraction is expected: for $\langle N_{\rm cen}\rangle<1$
halos, the central-satellite independent case has a lower number of
central-satellite pairs per halo compared to our fiducial case, which would
predict weaker small-scale clustering; the model compensates this by having a
higher $\langle N_{\rm sat}\rangle$. The $\chi^2$ values are, however,
generally similar to the previous results, which implies that the current
data are not able to put strong constraints on the correlation between
central and satellite galaxies.

The exact level of the correlation between central and satellite galaxies is
determined by galaxy formation physics. As discussed in \cite{Zentner13}, it
may be related to the formation history of the dark matter halos, exhibiting
assembly bias \citep[e.g.][]{Gao05}, and it may be further enhanced by the
phenomenon of ``galactic conformity'' \citep[e.g.][]{Weinmann06}. For the
galaxy samples considered in this work, the host halo masses are about two
orders of magnitude higher than the non-linear mass scale at $z\sim 0.5$, and
the halo assembly bias for such massive halos is expected to be small
\citep[e.g.][]{Wechsler06,Jing07}. In any case, our simple exercise here
gives us some idea on the magnitude of the uncertainty in the modeling
results (e.g. the satellite fraction) caused by potential galaxy/halo
assembly bias.

\subsection{Evolution of Satellite Galaxies}

HOD modeling results at different redshifts can be used to study galaxy
evolution \citep[e.g.][]{Zheng07,White07}. In order to study the evolution of
the LRGs in CMASS, we follow \cite{Guo13} and consider samples of constant
space number density from high to low redshifts, which allow for more direct
comparison with predictions of passive evolution. We construct two samples
with $n(z)=0.4\times 10^{-4} \,h^3\rm{Mpc}^{-3}$ (denoted as `low $n(z)$'
sample) and $n(z)=1.2\times 10^{-4} \,h^3\rm{Mpc}^{-3}$ (denoted as `moderate
$n(z)$' sample), respectively, using redshift-dependent luminosity
thresholds. We consider the evolution between two redshift ranges of
$0.566<z<0.616$ ($\bar{z}=0.591$) and $0.487<z<0.537$ ($\bar{z}=0.512$). The
redshifts are selected to ensure the completeness of the LRG samples as well
as to match the simulation outputs (see below). The median redshifts of
galaxies in these two redshift ranges are only slightly different from the
simulation outputs since we are considering narrow redshift intervals of
$\Delta z=0.05$, therefore the choices of the redshift ranges do not affect
our conclusions.

In \citet{Guo13}, based on the evolution of large-scale bias factors, we
showed that the evolution of fixed number-density samples is consistent with
passive evolution \citep{Fry96}. Such a simple model is not sensitive to the
evolution of satellites. Here we follow a similar method as in \cite{White07}
(see also \citealt*{Seo08}) to study the evolution of satellite galaxies. For
each constant number density sample, we perform HOD modeling at the two
redshifts and infer the corresponding HODs. We then populate halos identified
at $z=0.591$ (high-$z$) in an $N$-body simulation based on the high-$z$ HOD
solutions, by using particles to represent galaxies. These particles are
tracked in the simulation to $z=0.512$ (low-$z$) to derive the passively
evolved HOD. In such a passive evolution, each galaxy keeps its own identity
and there is no merging and disruption. The difference between the passively
evolved HOD and the HOD inferred from the low-$z$ clustering then allows us
to study galaxy evolution.

We use a set of 16 simulations from \cite*{Jing07}, which employ $1024^3$
particles of mass $1.5\times10^{10}\msun$ in a periodic box of $600\mpchi$ on
a side (detailed in Jing et al. 2007). The cosmological parameters in the
simulations are $\Omega_m=0.268$, $\Omega_{\Lambda}=0.732$, $h=0.71$,
$\Omega_b$=0.045, and $\sigma_8=0.85$. The primordial density field is
assumed to be a Gaussian distribution with a scale-invariant power spectrum.
The linear power spectrum is described using the transfer function of
\cite{Eisenstein98}. To be consistent, we adopt this cosmology for the 2PCF
measurements and HOD modeling in this subsection. For each fixed number
density sample, we populate $z=0.591$ halos using the corresponding
best-fitting HOD and track the `galaxies' to $z=0.512$. We then compare these
passive evolution predictions with the best-fitting HOD model at this lower
redshift. The results for the mean satellite occupation functions are
presented in the top panels of Figure~\ref{fig:passive}. In each panel, the
solid curve is the best-fitting HOD model at $\bar{z}=0.512$ with the shaded
area showing the $1\sigma$ distribution around the best-fitting model. The
triangles are the passive evolution predictions from the simulation. The
evolution of the constant $n(z)$ samples generally agrees with the passive
evolution predictions, although the low-$n(z)$ sample shows a slight
deviation at the low-mass end.

To fully explore the evolution of the satellite galaxies, we also must take
into account the uncertainty in the high-redshift HOD models. For this
purpose, we randomly select $10,000$ models from the MCMC chains at $z=0.591$
and passively evolve them (by tracking the particles in the simulations) to
$z=0.512$. We then derive the mean satellite occupation numbers of the
$10,000$ models, which we denote as $\langle N_{\rm sat, passive}\rangle$, at
each halo mass. We also randomly select $10,000$ models from the MCMC chains
at $z=0.512$ and denote their mean satellite occupation number as $\langle
N_{\rm sat, HOD}\rangle$. We calculate from all of these models the
distribution of the ratio $\langle N_{\rm sat,passive}\rangle / \langle
N_{\rm sat, HOD}\rangle$, which is shown in the bottom panels of
Figure~\ref{fig:passive}, colour-coded for satellites in halos of different
masses. If galaxies only evolve passively, the ratio should be one. Given the
broad distribution of the $\langle N_{\rm sat,passive}\rangle / \langle
N_{\rm sat, HOD}\rangle$ ratio, our results are consistent with satellite
galaxies in both samples experiencing no substantial merging or disruption
during the above redshift interval.

\cite{White07} study the evolution of LRGs from $z{\sim}0.9$ to $z{\sim}0.5$
in the NOAO Deep Wide-Field Survey \cite[NDWFS;][]{Jannuzi99} and find that
about one third of the luminous satellite galaxies in massive halos appear to
undergo merging or disruption in this redshift range. The average
merger/disruption rate per Gyr is about $14$ per cent. Assuming the same
merger/disruption rate in our smaller redshift range, which spans about
0.5\,Gyr, the expected total merger/disruption rate of the satellite galaxies
would be about $7$ per cent. \cite{Wake08} also study the evolution of LRGs
from redshift $z=0.55$ to $z=0.19$, and find an average merger rate of $2.4$
per cent per $\rm{Gyr}$. Given the large uncertainty in the distributions
seen in the bottom panels of Figure~\ref{fig:passive}, our results are not
inconsistent with those in \cite{White07} and \cite{Wake08}. A larger
redshift interval would help to reduce the statistical uncertainties and
allow better constraints on the evolution of galaxies, which we will consider
in future work.

\section{Conclusion}\label{sec:conclusions}

In this paper, we perform HOD modeling of projected 2PCFs of CMASS galaxies
in the SDSS-III BOSS DR10, focusing on the dependence on galaxy colour and
luminosity. We study the relation of galaxies to dark matter halos, interpret
the trends with colour and luminosity, and investigate the implications for
galaxy distribution inside halos and galaxy evolution.

The galaxy-halo relations from our three luminosity threshold samples show
trends consistent with those in the SDSS MAIN and LRG samples
\citep{Zehavi05,Zehavi11,Zheng09}. The tight correlation between galaxy
luminosity and halo mass persists for luminous galaxies in massive halos at
$z\sim 0.5$. More luminous galaxies occupy more massive halos, with most of
galaxies in our samples residing in halos of $10^{13}$--$10^{14}\msun$ as
central galaxies. The fraction of satellite galaxies decreases with galaxy
luminosity threshold, varying from $\sim 8$ per cent for $M_i<-21.6$ galaxies
to $\sim 5$ per cent for $M_i<-22.0$ galaxies. Most of the satellite galaxies
reside in halos of mass $\sim 10^{14}\msun$.

For the characteristic halo mass scales $M_{\rm min}$ and  $M_1$ for central
and satellite galaxies, respectively, the gap between them is smaller for
more luminous galaxies, again consistent with the findings for the SDSS MAIN
sample \citep{Zehavi05,Zehavi11}. The ratio $M_1/M_{\rm min}$ for CMASS
galaxies is about $8.7$, significantly smaller than the ${\sim}17$ ratio for
the fainter MAIN sample galaxies \citep{Zehavi11}, but similar to that for
LRGs \citep{Zheng09}. The smaller $M_1/M_{\rm min}$ ratio implies a more
recent accretion of luminous satellite galaxies in massive halos.

For the three colour subsamples studied, in the luminosity range of
$-22.2<M_i<-21.6$ and redshift range of $0.48<z<0.55$, redder galaxies
exhibit stronger clustering amplitudes and steeper slopes of the projected
2PCFs on small scales,  while the large-scale clustering is similar. We
interpret the colour trend with a simple HOD model where the three samples
all have their central galaxies in halos of the same mass range (around
$10^{13}\msun$), but the satellite fraction is higher for redder galaxies
(see \citealt{Zehavi11}). A higher satellite fraction for redder galaxies
enhances the contribution from small-scale pairs, resulting in stronger
small-scale clustering. On large scales, the 2PCF is dominated by the central
galaxies contribution, and the similar halo mass scales for central galaxies
therefore lead to similar large-scale clustering amplitudes.

The accurate fibre-collision correction (Guo et al. 2012) enables a
measurement of the 2PCFs on small scales (below $\sim 0.4\mpchi$), providing
an opportunity to constrain the small-scale distribution of galaxies within
halos. We extend our modeling by replacing the NFW profile for satellite
galaxy distribution with a generalized profile, with two additional free
parameters. The NFW profile still provides a sufficient interpretation to the
small-scale clustering measurements of the luminosity-threshold samples. For
the colour subsamples, the `{\it reddest}' one favors a steeper profile on
small scales (close to an SIS profile), to match the steep rise of the 2PCF
below $\sim 0.2\mpchi$.

We attempt to study the evolution of CMASS galaxies inferred from HOD
modeling of the 2PCFs at two redshifts, $z=0.591$ and $z=0.512$, for samples
with constant number density. The HOD inferred from the clustering
measurements at the lower redshift is consistent with the one passively
evolved from the higher redshift, i.e. we do not find evidence for the
merging and disruption of luminous satellite galaxies during the above narrow
redshift interval. However, our resulting wide range of accepted models
provides little constraining power on the merging and disruption of satellite
galaxies. A larger redshift interval and high signal-to-noise ratio
clustering measurement would help to better study galaxy evolution with such
a method.

Measurements and modeling of small-scale redshift-space distortions of the
CMASS sample can provide additional important constraints on the spatial and
velocity distributions of galaxies inside halos, and will be presented
elsewhere in a forthcoming paper. The final SDSS-III BOSS DR12, which will be
available in late 2014, will present a $\sim40$ per cent increase over the
current sample analysed here. Improved measurements and modeling of the full
BOSS sample will provide stronger constraints and will greatly enhance our
understanding of the distribution and evolution of these massive galaxies.

\section*{Acknowledgments}

We thank Yipeng Jing for kindly providing the simulations used in this paper.
We thank Douglas F. Watson for helpful discussions and the anonymous referee
for useful comments which improved the presentation of this paper. H.G., Z.Z.
and I.Z. were supported by NSF grant AST-0907947. Z.Z. was partially
supported by NSF grant AST-1208891.

Funding for SDSS-III has been provided by the Alfred P. Sloan Foundation, the
Participating Institutions, the National Science Foundation, and the U.S.
Department of Energy Office of Science. The SDSS-III web site is
http://www.sdss3.org/.

SDSS-III is managed by the Astrophysical Research Consortium for the
Participating Institutions of the SDSS-III Collaboration including the
University of Arizona, the Brazilian Participation Group, Brookhaven National
Laboratory, University of Cambridge, Carnegie Mellon University, University
of Florida, the French Participation Group, the German Participation Group,
Harvard University, the Instituto de Astrofisica de Canarias, the Michigan
State/Notre Dame/JINA Participation Group, Johns Hopkins University, Lawrence
Berkeley National Laboratory, Max Planck Institute for Astrophysics, Max
Planck Institute for Extraterrestrial Physics, New Mexico State University,
New York University, Ohio State University, Pennsylvania State University,
University of Portsmouth, Princeton University, the Spanish Participation
Group, University of Tokyo, University of Utah, Vanderbilt University,
University of Virginia, University of Washington, and Yale University.

\begin{appendix}
\section{Tests of jackknife covariance matrix for HOD modeling}\label{app:jack}
\begin{figure*}
\includegraphics[width=0.9\textwidth]{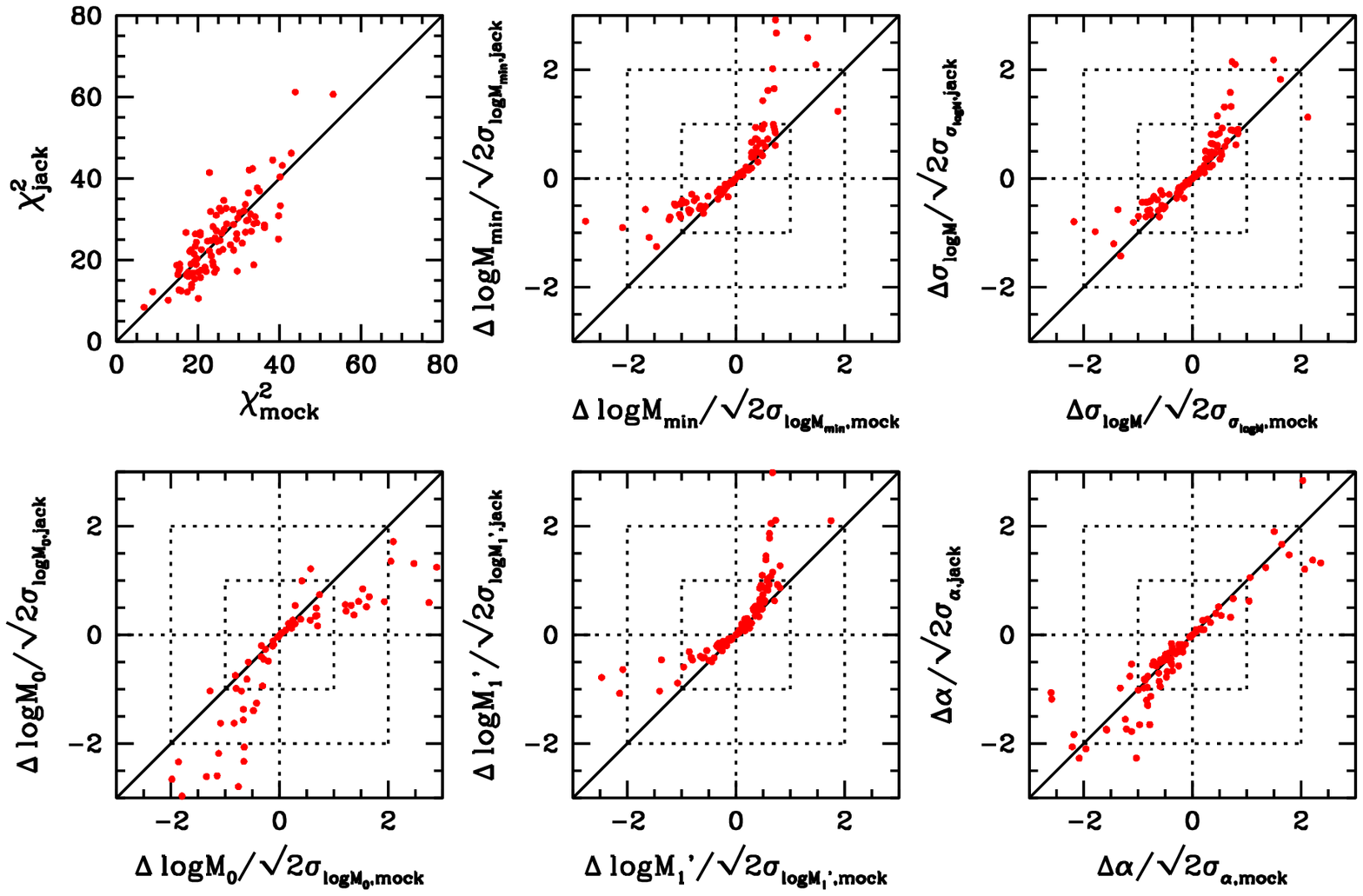}\caption{Comparison of best-fitting HOD parameters using mock and
jackknife covariance matrices. Top left panel shows the comparison of
$\chi^2$ for the best-fitting models using the two covariance matrices. The
comparisons for the five HOD parameters are presented in other panels, as
labelled. The red dots in each panel are the corresponding values for the
$100$ mock catalogues. The dotted boxes indicate the ranges where the
relative difference for each HOD parameter $\Delta p/\sqrt{2}\sigma_p=1$ or
$2$.} \label{fig:hodmockjack}
\end{figure*}
In large-scale clustering analyses, it has become customary to use large sets
of realistic mock catalogues, matching the observed clustering, to derive the
error covariance matrix. However, when studying small-scale clustering and in
particular when utilizing many subsamples with different clustering
properties, jackknife resampling is a far more practical tool, and has been
widely used for such studies (e.g. \citealt{Zehavi05,Zehavi11,Guo13}).

\cite{Guo13} demonstrated the accuracy of the jackknife errors for the
projected 2PCF $w_p(r_p)$ in comparison to the mock errors using a set of 100
mock catalogues from \cite{Manera13}. The ensemble average of jackknife
errors shows good agreement with the mock errors (see Appendix B and
specifically Figure~16 in \citealt{Guo13}). For a given realization/sample,
however, the jackknife covariance matrix still differs from the ensemble
average and the mock covariance matrix. Here, we specifically study the
accuracy of the jackknife covariance matrices when constraining the HOD
parameters and show the jackknife covariance matrix from an individual sample
works well in constraining the best-fitting HOD parameters.

We use the same set of 100 mock catalogues as in \cite{Guo13}. In each mock,
we measure $w_p(r_p)$ and the corresponding jackknife covariance matrix. We
define the mock covariance matrix as the variance of $w_p(r_p)$ among the
$100$ mock catalogues. For each mock, we determine the best-fitting HOD
parameters for the measured $w_p(r_p)$ in this mock using the corresponding
jackknife covariance matrix and compare to those parameters obtained from
using the mock covariance matrix. To have a fair comparison of the
best-fitting HOD parameters using the two covariance matrices, we define the
relative difference for each parameter $p$ as $\Delta p/\sqrt{2}\sigma_p$,
where $\Delta p\equiv p_{\rm mock}-p_{jack}$ is the difference between two
best-fitting parameters and $\sigma_{p,{\rm mock}/{\rm jack}}$ is the
marginalized 1$\sigma$ uncertainty on the parameter $p$ from modeling with
either the mock or jackknife covariance matrix. The factor $\sqrt{2}$ comes
from the fact that we are comparing the differences between two parameters,
$p_{\rm mock}$ and $p_{\rm jack}$.

Figure~\ref{fig:hodmockjack} displays the comparison of $\chi^2$ for the
best-fitting models using the two covariance matrices, as well as the
relative differences for the five HOD parameters. The red dots in each panel
are the corresponding values for the $100$ mock catalogues. The degree of
freedom (dof) of the fitting is $14+1-5=10$. It is clear that the $\chi^2$
values of using the jackknife and mock covariance matrices are quite similar,
implying that the two covariance matrices would produce similar
goodness-of-fit for each mock. The large $\chi^2/\rm{dof}$ values could be
caused by the fact that the mock catalogues do not fully describe the
realistic galaxy clustering on small scales that can be captured by our HOD
model \citep*[see e.g.][]{White14}.

As shown in Figure~\ref{fig:hodmockjack}, about $68$ per cent of the
best-fitting HOD parameters derived from the two covariance matrices are
within $1\sigma$ range of each other and about $92$ per cent lie within the
$2\sigma$ errors. The parameter $\log M_0$, which determines the cut-off mass
scale of the satellite galaxies and has more freedom in the models, is less
constrained compared to other parameters. It is evident from the figure that
the errors on the best-fitting HOD parameters derived from the two covariance
matrices are also quite similar. We thus conclude that the jackknife
covariance matrices will provide reasonably good estimates for best-fitting
HOD parameters, compared to using the mock covariance matrices.

\section{Correlation Function Measurements}\label{app:wp}
We present the measurements of the projected 2PCFs $w_p(r_p)$ together with
the diagonal errors used in this paper for the three luminosity-threshold
samples and three colour samples in Table~\ref{tab:wprplum}.
\begin{table*}
\begin{minipage}{166mm}
\caption{Measurements of $w_p(r_p)$ for the luminosity-threshold and finer
colour samples} \label{tab:wprplum}
\begin{tabular}{@{}lrrrrrr@{}}
\hline
$r_p$    &$M_i<-21.6$ & $M_i<-21.8$ & $M_i<-22.0$ & `{\it green}' & `{\it redseq}' & `{\it reddest}'\\
\hline
0.021    &10029.85 (7420.09) &18138.15 (11600.27)&6986.76 (5854.06)&---\quad\qquad\qquad&5079.46 (4095.05) &25888.67
(12384.27)\\
0.033    &4744.70 (883.75)&6285.53 (1823.47)& 12368.79 (4577.34)&---\quad\qquad\qquad& 3837.92 (2098.00) & 12611.48
(3889.88)\\
0.052    &2860.37 (359.76)&4659.69 (769.40)&5602.46 (1270.36)&1139.96 (782.25)&4282.92 (1368.53)&7798.45 (1785.71)\\
0.082    &2732.31 (236.19)&3968.58 (420.99)&5185.31 (911.10)&2386.38 (849.47)&1772.59 (592.95)&3010.72 (722.82)\\
0.129    &1560.89 (105.03)& 1774.97 (171.64)& 2225.19 (363.07)&496.75 (219.62)& 893.92 (266.46)& 3015.35 (414.65)\\
0.205    &1025.62 (60.78)& 1211.53 (108.33)& 1792.21 (245.61)& 562.86 (153.35)& 820.90 (160.24)& 1312.50 (200.97)\\
0.325    & 629.37 (30.65)& 830.34 (54.38)& 1009.84 (106.53)& 370.83 (95.63)& 591.31 (103.87)& 723.29 (89.62)\\
0.515    & 363.88 (12.43)& 441.23 (19.88)& 620.87 (43.63)& 189.24 (34.62)& 333.02 (33.20)& 456.22 (37.48)\\
0.815    & 195.87 (6.92)& 233.84 (12.04)& 333.99 (23.18)& 149.75 (21.26)& 178.01 (20.85)& 226.14 (20.98)\\
1.292    & 128.80 (4.58)& 154.92 (6.84)& 199.46 (14.81)& 97.87 (14.13)& 124.07 (13.08)& 136.52 (11.92)\\
2.048    & 92.67 (2.99)& 106.21 (4.35)& 124.62 (7.34)& 70.30 (7.27)& 87.86 (7.62)& 112.16 (7.64)\\
3.246    & 67.65 (2.09)& 80.95 (3.03)& 101.29 (5.62)& 55.93 (5.53)& 53.45 (5.13)& 71.87 (4.97)\\
5.145    & 48.11 (1.55)& 55.99 (2.47)& 65.71 (4.36)& 39.00 (3.61)& 45.81 (3.24)& 56.05 (3.14)\\
8.155    & 32.13 (1.33)& 36.70 (1.78)& 44.27 (2.58)& 28.13 (2.48)& 29.49 (2.44)& 37.54 (2.29)\\
12.92    & 19.56 (1.13)& 22.27 (1.43)& 24.94 (2.11)& 14.88 (1.68)& 18.57 (1.77)& 24.36 (1.89)\\
20.48    & 10.59 (0.87)& 12.80 (1.12)& 15.17 (1.54)& 7.99 (1.31)& 10.89 (1.26)& 11.16 (1.19)\\
32.46    & 3.73 (0.62)& 4.58 (0.85)& 5.82 (1.21)& 2.86 (0.85)& 3.85 (0.93)& 4.75 (0.92)\\
51.45    & 1.24 (0.50)& 2.03 (0.66)& 2.64 (0.91)& 1.38 (0.64)& 1.09 (0.71)& 0.95 (0.75)\\
\hline
\end{tabular}
\medskip
The first column is the projected separation, $r_p$, in units of $\mpchi$.
The subsequent columns present the projected 2PCFs, $w_p(r_p)$, for different
samples. The diagonal errors are given in parentheses.
\end{minipage}
\end{table*}

\end{appendix}


\vspace{5pt}
\begin{thebibliography}{}
\bibitem[\protect\citeauthoryear{Ahn et al.}{2014}]{Ahn14} Ahn C.~P., et al.,
    2014, ApJS, 211, 17

\bibitem[\protect\citeauthoryear{Akaike}{1974}]{Akaike74} Akaike H., 1974,
    ITAC, 19, 716

\bibitem[\protect\citeauthoryear{Anderson et al.}{2012}]{Anderson12} Anderson
    L., et al., 2012, MNRAS, 427, 3435

\bibitem[\protect\citeauthoryear{Anderson et al.}{2013}]{Anderson13} Anderson
    L., et al., 2013, MNRAS submitted, arXiv:1312.4877

\bibitem[\protect\citeauthoryear{Bahcall \& Kulier}{2013}]{Bahcall13} Bahcall
    N.~A., Kulier A., 2013, MNRAS in press, arXiv:1310.0022

\bibitem[\protect\citeauthoryear{Benoist et al.}{1996}]{Benoist96} Benoist
    C., Maurogordato S., da Costa L.~N., Cappi A., Schaeffer R., 1996, ApJ,
    472, 452

\bibitem[\protect\citeauthoryear{Berlind \& Weinberg}{2002}]{Berlind02}
    Berlind A.~A., Weinberg D.~H., 2002, ApJ, 575, 587

\bibitem[\protect\citeauthoryear{Berlind et al.}{2003}]{Berlind03} Berlind
    A.~A., et al., 2003, ApJ, 593, 1

\bibitem[\protect\citeauthoryear{Beutler et al.}{2013}]{Beutler13} Beutler
    F., et al., 2013, MNRAS, 429, 3604

\bibitem[\protect\citeauthoryear{Blake, Collister, \& Lahav}{2008}]{Blake08}
    Blake C., Collister A., Lahav O., 2008, MNRAS, 385, 1257

\bibitem[\protect\citeauthoryear{Bolton et al.}{2012}]{Bolton12} Bolton
    A.~S., et al., 2012, AJ, 144, 144

\bibitem[\protect\citeauthoryear{Brown et al.}{2008}]{Brown08} Brown
    M.~J.~I., et al., 2008, ApJ, 682, 937

\bibitem[\protect\citeauthoryear{Budav{\'a}ri et al.}{2003}]{Budavari03}
    Budav{\'a}ri T., et al., 2003, ApJ, 595, 59

\bibitem[\protect\citeauthoryear{Bullock et al.}{2001}]{Bullock01} Bullock
    J.~S., Kolatt T.~S., Sigad Y., Somerville R.~S., Kravtsov A.~V., Klypin
    A.~A., Primack J.~R., Dekel A., 2001, MNRAS, 321, 559

\bibitem[\protect\citeauthoryear{Christodoulou et
    al.}{2012}]{Christodoulou12} Christodoulou L., et al., 2012, MNRAS, 425,
    1527

\bibitem[\protect\citeauthoryear{Coil et al.}{2006}]{Coil06} Coil A.~L.,
    Newman J.~A., Cooper M.~C., Davis M., Faber S.~M., Koo D.~C., Willmer
    C.~N.~A., 2006, ApJ, 644, 671

\bibitem[\protect\citeauthoryear{Coil et al.}{2008}]{Coil08} Coil A.~L., et
    al., 2008, ApJ, 672, 153

\bibitem[\protect\citeauthoryear{Coupon et al.}{2012}]{Coupon12} Coupon J.,
    et al., 2012, A\&A, 542, A5

\bibitem[\protect\citeauthoryear{Davis \& Geller}{1976}]{Davis76} Davis M.,
    Geller M.~J., 1976, ApJ, 208, 13

\bibitem[\protect\citeauthoryear{Davis et al.}{1988}]{Davis88} Davis M.,
    Meiksin A., Strauss M.~A., da Costa L.~N., Yahil A., 1988, ApJ, 333, L9

\bibitem[\protect\citeauthoryear{Dawson et al.}{2013}]{Dawson13} Dawson
    K.~S., et al., 2013, AJ, 145, 10

\bibitem[\protect\citeauthoryear{Eisenstein \& Hu}{1998}]{Eisenstein98}
    Eisenstein D.~J., Hu W., 1998, ApJ, 496, 605

\bibitem[\protect\citeauthoryear{Eisenstein et al.}{2011}]{Eisenstein11}
    Eisenstein D.~J., et al., 2011, AJ, 142, 72

\bibitem[\protect\citeauthoryear{Fry}{1996}]{Fry96} Fry J.~N., 1996, ApJ,
    461, L65

\bibitem[\protect\citeauthoryear{Fukugita et al.}{1996}]{Fukugita96} Fukugita
    M., Ichikawa T., Gunn J.~E., Doi M., Shimasaku K., Schneider D.~P., 1996,
    AJ, 111, 1748

\bibitem[Gao et al.(2005)]{Gao05} Gao, L., Springel, V., \& White, S.~D.~M.\
    2005, MNRAS, 363, L66

\bibitem[\protect\citeauthoryear{Grillo}{2012}]{Grillo12} Grillo C., 2012,
    ApJ, 747, L15

\bibitem[\protect\citeauthoryear{Gunn et al.}{1998}]{Gunn98} Gunn J.~E., et
    al., 1998, AJ, 116, 3040

\bibitem[\protect\citeauthoryear{Gunn et al.}{2006}]{Gunn06} Gunn J.~E., et
    al., 2006, AJ, 131, 2332

\bibitem[\protect\citeauthoryear{Guo et al.}{2014}]{Guo14} Guo H., Li C.,
    Jing Y.~P., B{\"o}rner G., 2014, ApJ, 780, 139

\bibitem[\protect\citeauthoryear{Guo, Zehavi, \& Zheng}{2012}]{Guo12} Guo H.,
    Zehavi I., Zheng Z., 2012, ApJ, 756, 127

\bibitem[\protect\citeauthoryear{Guo et al.}{2013}]{Guo13} Guo H., et al.,
    2013, ApJ, 767, 122 [G13]

\bibitem[\protect\citeauthoryear{Guzzo et al.}{1997}]{Guzzo97} Guzzo L.,
    Strauss M.~A., Fisher K.~B., Giovanelli R., Haynes M.~P., 1997, ApJ, 489,
    37

\bibitem[\protect\citeauthoryear{Hamilton}{1988}]{Hamilton88} Hamilton
    A.~J.~S., 1988, ApJ, 331, L59

\bibitem[\protect\citeauthoryear{Hartlap, Simon, \&
    Schneider}{2007}]{Hartlap07} Hartlap J., Simon P., Schneider P., 2007,
    A\&A, 464, 399

\bibitem[\protect\citeauthoryear{Jannuzi \& Dey}{1999}]{Jannuzi99} Jannuzi
    B.~T., Dey A., 1999, ASPC, 191, 111

\bibitem[\protect\citeauthoryear{Jiang, Hogg, \& Blanton}{2012}]{Jiang12}
    Jiang T., Hogg D.~W., Blanton M.~R., 2012, ApJ, 759, 140

\bibitem[\protect\citeauthoryear{Jing, Mo, \& Boerner}{1998}]{Jing98} Jing
    Y.~P., Mo H.~J., Boerner G., 1998, ApJ, 494, 1

\bibitem[\protect\citeauthoryear{Jing, Suto, \& Mo}{2007}]{Jing07} Jing
    Y.~P., Suto Y., Mo H.~J., 2007, ApJ, 657, 664

\bibitem[\protect\citeauthoryear{Kaiser}{1987}]{Kaiser87} Kaiser N., 1987,
    MNRAS, 227, 1

\bibitem[\protect\citeauthoryear{Kravtsov et al.}{2004}]{Kravtsov04} Kravtsov
    A.~V., Berlind A.~A., Wechsler R.~H., Klypin A.~A., Gottl{\"o}ber S.,
    Allgood B., Primack J.~R., 2004, ApJ, 609, 35

\bibitem[\protect\citeauthoryear{Kulkarni et al.}{2007}]{Kulkarni07} Kulkarni
    G.~V., Nichol R.~C., Sheth R.~K., Seo H.-J., Eisenstein D.~J., Gray A.,
    2007, MNRAS, 378, 1196

\bibitem[\protect\citeauthoryear{Landy \& Szalay}{1993}]{Landy93} Landy
    S.~D., Szalay A.~S., 1993, ApJ, 412, 64

\bibitem[\protect\citeauthoryear{Li et al.}{2006}]{Li06} Li C., Kauffmann G.,
    Jing Y.~P., White S.~D.~M., B{\"o}rner G., Cheng F.~Z., 2006, MNRAS, 368,
    21

\bibitem[\protect\citeauthoryear{Loh et al.}{2010}]{Loh10} Loh Y.-S., et al.,
    2010, MNRAS, 407, 55

\bibitem[\protect\citeauthoryear{Loveday et al.}{1995}]{Loveday95} Loveday
    J., Maddox S.~J., Efstathiou G., Peterson B.~A., 1995, ApJ, 442, 457

\bibitem[\protect\citeauthoryear{Madgwick et al.}{2003}]{Madgwick03} Madgwick
    D.~S., et al., 2003, MNRAS, 344, 847

\bibitem[\protect\citeauthoryear{Mandelbaum et al.}{2006}]{Mandelbaum06}
    Mandelbaum R., Seljak U., Kauffmann G., Hirata C.~M., Brinkmann J., 2006,
    MNRAS, 368, 715

\bibitem[\protect\citeauthoryear{Manera et al.}{2013}]{Manera13} Manera M.,
    et al., 2013, MNRAS, 428, 1036

\bibitem[\protect\citeauthoryear{Masjedi et al.}{2006}]{Masjedi06} Masjedi
    M., et al., 2006, ApJ, 644, 54

\bibitem[\protect\citeauthoryear{Meneux et al.}{2006}]{Meneux06} Meneux B.,
    et al., 2006, A\&A, 452, 387

\bibitem[\protect\citeauthoryear{Meneux et al.}{2008}]{Meneux08} Meneux B.,
    et al., 2008, A\&A, 478, 299

\bibitem[\protect\citeauthoryear{Meneux et al.}{2009}]{Meneux09} Meneux B.,
    et al., 2009, A\&A, 505, 463

\bibitem[\protect\citeauthoryear{Miyatake et al.}{2013}]{Miyatake13}
    Miyatake H., et al., 2013, ApJ submitted, arXiv:1311.1480

\bibitem[\protect\citeauthoryear{Navarro, Frenk, \& White}{1997}]{Navarro97}
    Navarro J.~F., Frenk C.~S., White S.~D.~M., 1997, ApJ, 490, 493

\bibitem[\protect\citeauthoryear{Norberg et al.}{2001}]{Norberg01} Norberg
    P., et al., 2001, MNRAS, 328, 64

\bibitem[\protect\citeauthoryear{Norberg et al.}{2002}]{Norberg02} Norberg
    P., et al., 2002, MNRAS, 332, 827

\bibitem[\protect\citeauthoryear{Padmanabhan et al.}{2009}]{Padmanabhan09}
    Padmanabhan N., White M., Norberg P., Porciani C., 2009, MNRAS, 397, 1862

\bibitem[\protect\citeauthoryear{Parejko et al.}{2013}]{Parejko13} Parejko
    J.~K., et al., 2013, MNRAS, 429, 98

\bibitem[\protect\citeauthoryear{Peacock \& Smith}{2000}]{Peacock00} Peacock
    J.~A., Smith R.~E., 2000, MNRAS, 318, 1144

\bibitem[\protect\citeauthoryear{Phleps et al.}{2006}]{Phleps06} Phleps S.,
    Peacock J.~A., Meisenheimer K., Wolf C., 2006, A\&A, 457, 145

\bibitem[\protect\citeauthoryear{Ross \& Brunner}{2009}]{Ross09} Ross A.~J.,
    Brunner R.~J., 2009, MNRAS, 399, 878

\bibitem[\protect\citeauthoryear{Ross et al.}{2011}]{Ross11} Ross A.~J., et
    al., 2011, MNRAS, 417, 1350

\bibitem[\protect\citeauthoryear{Ross, Percival, \& Brunner}{2010}]{Ross10}
    Ross A.~J., Percival W.~J., Brunner R.~J., 2010, MNRAS, 407, 420

\bibitem[\protect\citeauthoryear{Schlegel, Finkbeiner, \&
    Davis}{1998}]{Schlegel98} Schlegel D.~J., Finkbeiner D.~P., Davis M.,
    1998, ApJ, 500, 525

\bibitem[\protect\citeauthoryear{Scoccimarro et al.}{2001}]{Scoccimarro01}
    Scoccimarro R., Sheth R.~K., Hui L., Jain B., 2001, ApJ, 546, 20

\bibitem[\protect\citeauthoryear{Seljak}{2000}]{Seljak00} Seljak U., 2000,
    MNRAS, 318, 203

\bibitem[\protect\citeauthoryear{Seo, Eisenstein, \& Zehavi}{2008}]{Seo08}
    Seo H.-J., Eisenstein D.~J., Zehavi I., 2008, ApJ, 681, 998

\bibitem[\protect\citeauthoryear{Simon et al.}{2009}]{Simon09} Simon P.,
    Hetterscheidt M., Wolf C., Meisenheimer K., Hildebrandt H., Schneider P.,
    Schirmer M., Erben T., 2009, MNRAS, 398, 807

\bibitem[\protect\citeauthoryear{Skibba \& Sheth}{2009}]{Skibba09} Skibba
    R.~A., Sheth R.~K., 2009, MNRAS, 392, 1080

\bibitem[\protect\citeauthoryear{Skibba, Sheth, \& Martino}{2007}]{Skibba07}
    Skibba R.~A., Sheth R.~K., Martino M.~C., 2007, MNRAS, 382, 1940

\bibitem[\protect\citeauthoryear{Skibba et al.}{2013}]{Skibba13} Skibba
    R.~A., et al., 2013, arXiv, arXiv:1310.1093

\bibitem[\protect\citeauthoryear{Smee et al.}{2013}]{Smee13} Smee S.~A., et
    al., 2013, AJ, 146, 32

\bibitem[\protect\citeauthoryear{Swanson et al.}{2008}]{Swanson08} Swanson
    M.~E.~C., Tegmark M., Blanton M., Zehavi I., 2008, MNRAS, 385, 1635

\bibitem[\protect\citeauthoryear{Tinker et al.}{2005}]{Tinker05} Tinker
    J.~L., Weinberg D.~H., Zheng Z., Zehavi I., 2005, ApJ, 631, 41

\bibitem[\protect\citeauthoryear{Tojeiro et al.}{2012}]{Tojeiro12} Tojeiro
    R., et al., 2012, MNRAS, 424, 136

\bibitem[\protect\citeauthoryear{van den Bosch et al.}{2013}]{Bosch13} van
    den Bosch F.~C., More S., Cacciato M., Mo H., Yang X., 2013, MNRAS, 430,
    725

\bibitem[\protect\citeauthoryear{Wake et al.}{2008}]{Wake08} Wake D.~A., et
    al., 2008, MNRAS, 387, 1045

\bibitem[\protect\citeauthoryear{Wake et al.}{2011}]{Wake11} Wake D.~A., et
    al., 2011, ApJ, 728, 46

\bibitem[\protect\citeauthoryear{Wang et al.}{2007}]{Wang07} Wang Y., Yang
    X., Mo H.~J., van den Bosch F.~C., 2007, ApJ, 664, 608

\bibitem[\protect\citeauthoryear{Watson et al.}{2012}]{Watson12} Watson
    D.~F., Berlind A.~A., McBride C.~K., Hogg D.~W., Jiang T., 2012, ApJ,
    749, 83

\bibitem[\protect\citeauthoryear{Watson et al.}{2010}]{Watson10} Watson
    D.~F., Berlind A.~A., McBride C.~K., Masjedi M., 2010, ApJ, 709, 115

\bibitem[Wechsler et al.(2006)]{Wechsler06} Wechsler, R.~H., Zentner, A.~R.,
    Bullock, J.~S., Kravtsov, A.~V., \& Allgood, B.\ 2006, ApJ, 652, 71

\bibitem[Weinmann et al.(2006)]{Weinmann06} Weinmann, S.~M., van den Bosch,
    F.~C., Yang, X., \& Mo, H.~J.\ 2006, MNRAS, 366, 2

\bibitem[\protect\citeauthoryear{White et al.}{2011}]{White11} White M., et
    al., 2011, ApJ, 728, 126

\bibitem[\protect\citeauthoryear{White, Tinker, \& McBride}{2014}]{White14}
    White M., Tinker J.~L., McBride C.~K., 2014, MNRAS, 437, 2594

\bibitem[\protect\citeauthoryear{White et al.}{2007}]{White07} White M.,
    Zheng Z., Brown M.~J.~I., Dey A., Jannuzi B.~T., 2007, ApJ, 655, L69

\bibitem[\protect\citeauthoryear{Yang et al.}{2005}]{Yang05} Yang X., Mo
    H.~J., Jing Y.~P., van den Bosch F.~C., 2005, MNRAS, 358, 217

\bibitem[\protect\citeauthoryear{Yang, Mo, \& van den Bosch}{2003}]{Yang03}
    Yang X., Mo H.~J., van den Bosch F.~C., 2003, MNRAS, 339, 1057

\bibitem[\protect\citeauthoryear{York et al.}{2000}]{York00} York D.~G., et
    al., 2000, AJ, 120, 1579

\bibitem[\protect\citeauthoryear{Zehavi et al.}{2002}]{Zehavi02} Zehavi I.,
    et al., 2002, ApJ, 571, 172

\bibitem[\protect\citeauthoryear{Zehavi et al.}{2005}]{Zehavi05} Zehavi I.,
    et al., 2005b, ApJ, 630, 1

\bibitem[\protect\citeauthoryear{Zehavi et al.}{2011}]{Zehavi11} Zehavi I.,
    et al., 2011, ApJ, 736, 59

\bibitem[\protect\citeauthoryear{Zentner et al.}{2005}]{Zentner05} Zentner
    A.~R., Berlind A.~A., Bullock J.~S., Kravtsov A.~V., Wechsler R.~H.,
    2005, ApJ, 624, 505

\bibitem[\protect\citeauthoryear{Zentner, Hearin, \& van den
    Bosch}{2013}]{Zentner13} Zentner A.~R., Hearin A.~P., van den Bosch
    F.~C., 2013, MNRAS submitted, arXiv:1311.1818

\bibitem[\protect\citeauthoryear{Zhao et al.}{2009}]{Zhao09} Zhao D.~H., Jing
    Y.~P., Mo H.~J., B{\"o}rner G., 2009, ApJ, 707, 354

\bibitem[\protect\citeauthoryear{Zheng}{2004}]{Zheng04} Zheng Z., 2004, ApJ,
    610, 61

\bibitem[\protect\citeauthoryear{Zheng et al.}{2005}]{Zheng05} Zheng Z., et
    al., 2005, ApJ, 633, 791

\bibitem[\protect\citeauthoryear{Zheng, Coil, \& Zehavi}{2007}]{Zheng07}
    Zheng Z., Coil A.~L., Zehavi I., 2007, ApJ, 667, 760

\bibitem[\protect\citeauthoryear{Zheng et al.}{2002}]{Zheng02} Zheng Z.,
    Tinker J.~L., Weinberg D.~H., Berlind A.~A., 2002, ApJ, 575, 617

\bibitem[\protect\citeauthoryear{Zheng et al.}{2009}]{Zheng09} Zheng Z.,
    Zehavi I., Eisenstein D.~J., Weinberg D.~H., Jing Y.~P., 2009, ApJ, 707,
    554

\bibitem[\protect\citeauthoryear{Zu et al.}{2008}]{Zu08} Zu Y., Zheng Z., Zhu
    G., Jing Y.~P., 2008, ApJ, 686, 41

\end{thebibliography}
\end{document}